\documentclass[11pt,a4paper]{article}
\usepackage{amsfonts}
\usepackage{epsfig}
\usepackage{graphicx}
\usepackage{amsmath}
\usepackage{amssymb}
\usepackage{color}
\usepackage{cite}

\usepackage[colorlinks=true, linkcolor=blue, urlcolor=blue, citecolor=blue]{hyperref}

\begin{document}

\title{\bf Kinks in composite scalar field theories}

\author{A. Alonso-Izquierdo$^{(a,b)}$, A.J. Balseyro Sebastian$^{(b)}$ and\\  M. A. Gonzalez Leon$^{(a,b)}$
	\\ {\normalsize {\it $^{(a)}$ Departamento de Matematica
			Aplicada}, {\it Universidad de Salamanca, SPAIN}}\\ {\normalsize {\it $^{(b)}$ IUFFyM}, {\it Universidad de Salamanca, SPAIN}}\\\\
	Email:  
	\href{mailto:alonsoiz@usal.es}{alonsoiz@usal.es}, 
	\href{mailto:albalse@usal.es}{albalse@usal.es}, 
	\href{mailto:magleon@usal.es}{magleon@usal.es}
}

\date{}

\maketitle

\begin{abstract}

	In this work, families of kinks are analytically identified in multifield theories with either polynomial or deformed sine-Gordon-type potentials. The underlying procedure not only allows us to obtain analytical solutions for these models, but also provides a framework for constructing more general families of field theories that inherit certain analytical information about their solutions. Specifically, this method combines two known field theories into a new composite field theory whose target space is the product of the original target spaces. By suitably coupling the fields through a superpotential defined on the product space, the dynamics in the subspaces become entangled while preserving original kinks as boundary kinks. Different composite field theories are studied, including extensions of well-known models to wider target spaces.

\noindent

\end{abstract}

%%%%%%%%%%%%%%%%%%%%%%%%%%%%%%%%%%%%%%%%%%%%%%%%%%%%%%%%%%%%%%%%%%%%%%%%%%%%%%%%%%%%%%%%%%%%%%%%%%%%%%%%%%%%%%%%%%%%%%%%%

\section{Introduction}

Solitons, solutions of non-linear field theories with finite energy whose energy density travels at constant velocity and without losing its form, have been an active field of research during the past decades. Not only because of its interesting mathematical properties, but also because of its ubiquity in physics. Indeed, solitons are present in fields such as Condensed Matter \cite{Bishop1980, Cheng2021, Eschenfelder1981, Jona1993, Machon2016, Pollard2020, Strukov1998, Vanhaverbeke2008}, Molecular physics \cite{Bazeia1999, Cortijo2007, Davydov1985, Guillamat2022, Yakushevich2004}, Optics \cite{Ablowitz2022, Agrawall1995, Mollenauer2006, Schneider2004} and Cosmology and high-energy physics \cite{Kibble1976, Kolb1990, Vachaspati2006, Vilenkin1985, Vilenkin2000, Weinberg2012} among several others. Obtaining analytical solutions of non-linear differential equations is highly challenging in general in non-integrable systems. In this work we shall be focusing on solitons for field theories defined on the $1+1-$dimensional Minkowski space, known in the literature as kinks. Several techniques have been developed to assist in this task. One approach to generating new field theories with identifiable solitons is the so-called deformation method, in which solutions from one theory are transferred to another \cite{Afonso2007, Almeida2004, AlonsoIzquierdo2013,AlonsoIzquierdo2025,Bazeia2002, Bazeia2006a, Bazeia2006b, Bazeia2011, Bazeia2013, Bazeia2017, Chumbes2010, Cruz2009}.

In the present study analytical solutions are identified in a family of field theories with polynomial potentials. The analysis of the properties that these models exhibit allows us to identify a method for combining field theories while retaining some analytical information about solutions. In particular, given two models, families of potentials will be constructed on the product of the target spaces, leading to a coupled field theory that exhibits a richer variety of kinks. While solutions from the original models will be inherited in the new model, new ones will appear due to the coupling. This procedure will also be employed to construct extensions of known models, for which the kink variety can be explicitly identified. For instance, by combining sine-Gordon models, an extension in a $n-$dimensional target space of the sine-Gordon model with $n$ fields can be constructed. In this example, the deformation of the potential is obtained by means of a coupling that breaks the integrability of the model. This is of particular interest in all phenomena described by the sine-Gordon equation, as the new degree of freedom may help to predict experiments more accurately. For example, in the context of the Josephson effect with multiple coupled junction devices, perturbing the sine-Gordon equations leads to a description of fluxons in these junctions \cite{Yukon1991,Abdufarrukh2004}. Other examples where perturbing the sine-Gordon model may be useful are models for the angles of rotation of bases in DNA  \cite{Homma1984,Zdravković2013} or inflationary models in the early universe \cite{Freese1990}. Similarly, this method can be employed to extend other models like $\phi^4-$models with multiple fields. This may have interesting applications in contexts such as the theory of Ginzburg-Landau \cite{Ginzburg1950,Cyrot1973}, where terms of the form of the $\phi^4-$model are included to describe the behavior of superconductors, or in the context of inflatons in inflationary cosmology \cite{Linde1990}. 

The aim of this work is to explore specific types of combinations of field theories that preserve analytical information. The structure of this paper is as follows. In Section $2$ the type of field theories that we shall consider will be introduced and a very particular model with polynomial potential is studied. The properties of this model will allow us to generalize the procedure in Section $4$,  where other examples of ``composite'' field theories are constructed. In particular, two different extensions of the $\phi^4-$model to the plane and one extension of the sine-Gordon model with three fields shall be constructed. Finally, conclusions are drawn in Section $5$.   

\section{A scalar field theory with a tenth-order polynomial potential}

In this section, the type of field theories that shall be studied is presented. Given a configuration $\phi:\mathbb{R}^{1,1}\rightarrow \mathbb{R}^{n}$, let us consider a field theory for which the action can be written in Cartesian coordinates $\phi \equiv \left\{\phi^1,\dots,\phi^n\right\}$ as  
$$S\left[\phi\right]=\int_{\mathbb{R}^{1,1}}\left[\frac{1}{2} ~\displaystyle\sum_{i=1}^{n}\left[\left(\frac{d\phi^i}{dt}\right)^2-\left(\frac{d\phi^i}{dx}\right)^2\right]-V(\phi)\right] ~ dx dt \, ,$$
where $V$ is a non-negative potential function. This action leads to the Euler-Lagrange equations
\begin{equation}
	-\frac{\partial^2\phi^i}{\partial t^2}+\frac{\partial^2\phi^i}{\partial x^2}=\frac{\partial V}{\partial \phi^i}\, , \,  \, i=1,\dots, n \, , \label{eq:feSigmamodel21}
\end{equation}
where in general fields will be coupled. On the other hand, the energy for a given configuration $\phi:\mathbb{R}^{1,1}\rightarrow \mathbb{R}^{n}$ can be written as
\begin{equation*}
	E[\phi]= \int^{\infty}_{-\infty}  \, \left[ \frac{1}{2}\, \sum_{i=1}^{n}\left[\left(\frac{\partial\phi^i}{\partial t}\right)^2+\left(\frac{\partial\phi^i}{\partial x}\right)^2\right]+ V(\phi)\right] \, dx \,,
\end{equation*}
where the integrand, the energy density, will be denoted as $\varepsilon(x,t)$. In this work, we shall be interested in kinks, finite energy solutions of field theories in $\mathbb{R}^{1,1}$ whose energy density $\varepsilon(t,x)$ travels at constant velocity and preserving its form. The form of this functional forces all configurations in our finite energy configuration space to satisfy the asymptotic conditions:
\begin{equation*}
	\displaystyle\lim_{x\rightarrow\pm\infty} \frac{\partial \phi^i(t,x)}{\partial t}=\displaystyle\lim_{x\rightarrow\pm\infty}\frac{\partial \phi^i(t,x)}{\partial x}=0 \, , \quad \forall i \,, \qquad \displaystyle\lim_{x\rightarrow\pm\infty}\phi(t,x)\in \mathcal{M}\,,
\end{equation*} 
where $\mathcal{M}$ is the set of discrete vacua of the potential $V(\phi)$, \textit{i.e.} the points where it vanishes:
\begin{equation*}
	\mathcal{M}=\left\{\phi_ a\in \mathbb{R}^n \mid V(\phi_a)=0,\quad  a=1,2,\dots, \right\}\,,
\end{equation*}
  which is assumed to be discrete. This configuration space can then be decomposed into different topological sectors, each characterized by different asymptotic conditions. Given the Lorentz invariance of the action \eqref{eq:feSigmamodel21}, in order to identify kinks it will suffice to restrict the analysis to static configurations with finite energy. Moreover, if the potential $V(\phi)$ can be written in terms of a superpotential $W(\phi):\mathbb{R}^{n}\rightarrow\mathbb{R}$ as follows

\begin{equation}
	V(\phi)= \frac{1}{2}\sum_{i=1}^{n} \left(\frac{\partial W}{\partial \phi^i} \right)^2  ,  \label{ps1}
\end{equation}
 then the Bogomol'nyi arrangement \cite{Bogomolny1976} can be performed in the static energy functional. This allows us to find configurations that minimize the static energy in a given topological sector as solutions of the Bogomol'nyi equations
\begin{equation}
	\frac{d\phi^i}{dx}=(-1)^{\epsilon} \,  \frac{\partial W}{\partial \phi^i} \, ,\qquad i=1,\dots, n , \label{bpsb1}
\end{equation}
where $\epsilon=0,1$, which are also solutions of \eqref{eq:feSigmamodel21}. This also implies that the energy of these static configurations is a topological charge
\begin{equation}\label{eq:EnergyBPS1}
	T=\Big\vert \lim_{x\rightarrow \infty} W[\phi(x)]- \lim_{x\rightarrow -\infty}W[\phi(x)]\Big\vert \, .
\end{equation}

\begin{figure}[h!]
	\centering
	\includegraphics[height=4cm]{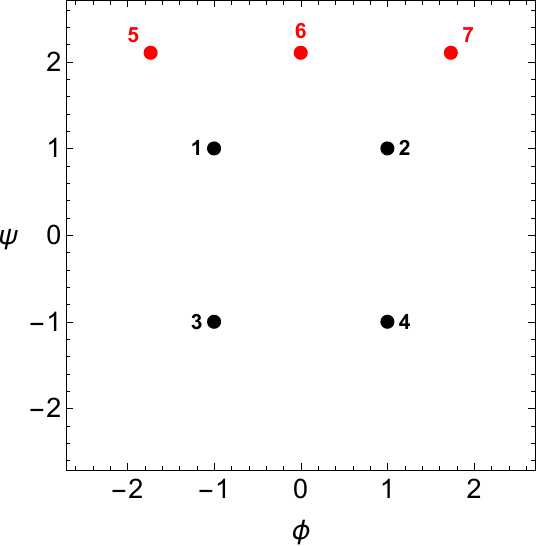}	\hspace{0.3cm}
	\includegraphics[height=4cm]{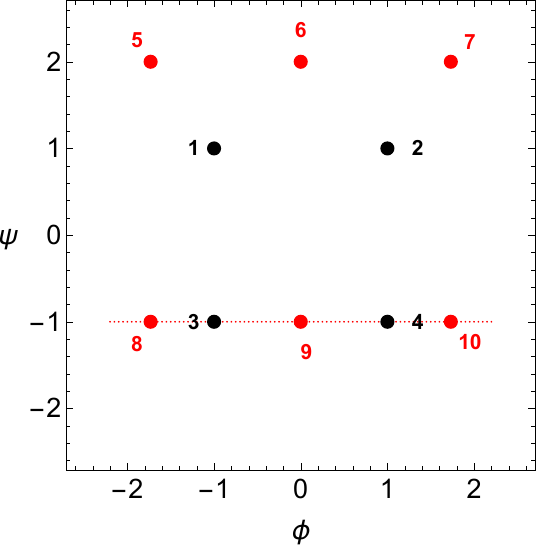} \hspace{0.3cm}
	\includegraphics[height=4cm]{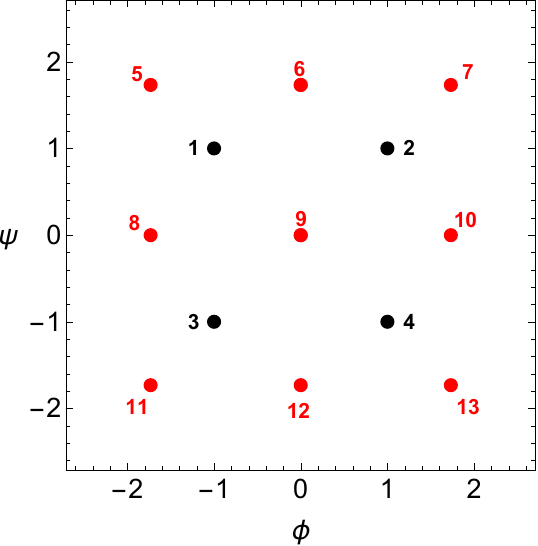}\\
	\includegraphics[height=4cm]{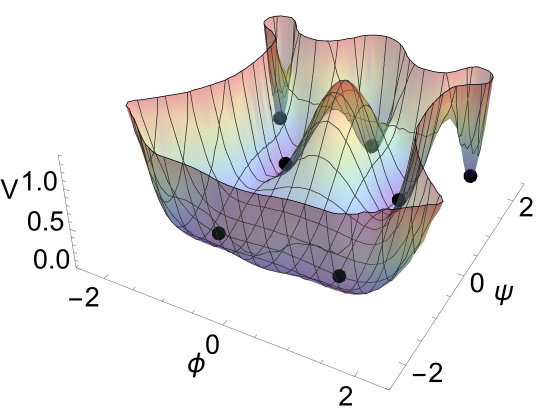}	\hspace{0.3cm}
	\includegraphics[height=4cm]{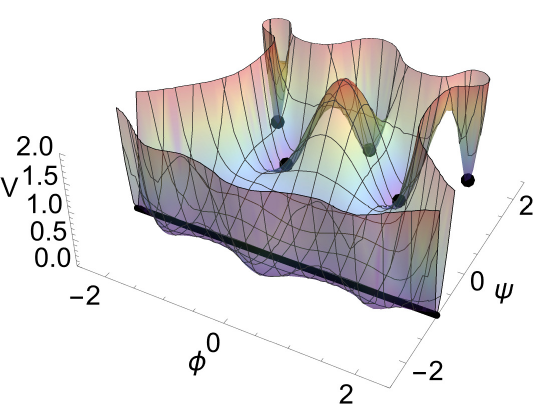} \hspace{0.3cm}
	\includegraphics[height=4cm]{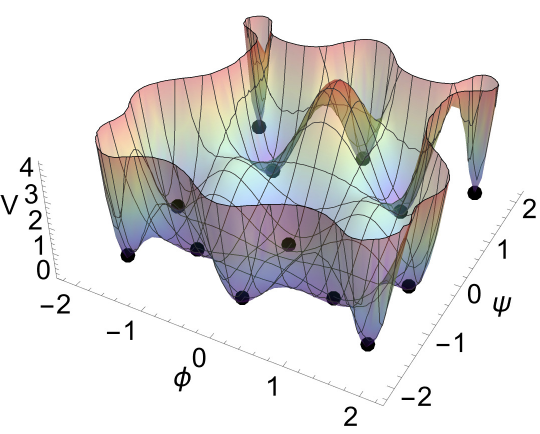}\\		
	\caption{\small Vacua of the field theory (top panels) and its corresponding potential (bottom panels) for $\sigma=1$ (left), $\sigma=\frac{3}{2}$ (middle) and $\sigma=3$ (right). In the upper panels, the four fixed vacua are represented in black while the variable ones are depicted in red.}
	\label{fig:CompositePhi4VacuaPotential}
\end{figure}

Let us now explore a specific field theory with two real scalar fields and a polynomial potential. This model is examined with the aim of gaining insight into certain properties of field theories that can give rise to analytical solutions. In particular, let us choose a one-parameter family of potential functions
\begin{equation}
	V(\phi,\psi;\sigma)=\frac{1}{2} \left[1+\sigma\left(\psi-\frac{\psi^3}{3}\right)\right]^2  (1-\phi^2)^2 + \frac{\sigma^2}{2} \left[ \left(\phi-\frac{\phi^3}{3}\right)\right]^2  (1-\psi^2)^2 \, ,\label{eq:PrimerPotencial}
\end{equation}
 where $\sigma\in \mathbb{R}$ is the parameter that labels each member of the family, and fields have been denoted as $(\phi^1,\phi^2)\rightarrow(\phi,\psi)$ to alleviate notation. Let us restrict the analysis to $\sigma>0$, as the cases with negative $\sigma$ follow directly from the relation $V(\phi,\psi,-\sigma)=V(\phi,-\psi,\sigma)$. Notice that the set of points where the second term of the potential vanishes is independent of $\sigma$, whereas that of the first term depends on this parameter. When $\sigma=0$ the potential simplifies enormously, becoming just the potential of a $\phi^4-$model with field $\phi$. For non-vanishing $\sigma$ and $\sigma\neq \frac{3}{2}$, the vacuum manifold is composed of four ``fixed'' points and ``variable'' points
\begin{equation*}
	\mathcal{M}_{\sigma}=\left\lbrace (-1,-1),(-1,1),(1,-1),(1,1),(0,p_j),(\sqrt{3},p_j),(-\sqrt{3},p_j) \right\rbrace \, ,
\end{equation*}
whose number and position are determined by $p_j$, the real roots of
\begin{equation}\label{eq:cubic}
	1+\sigma\left(p-\frac{p^3}{3}\right)=0 \,.
\end{equation}
 Notice how, however, when $\sigma=\frac{3}{2}$ a continuous vacua appear at $\psi=- 1$ respectively, see Figure \ref{fig:CompositePhi4VacuaPotential}. In the upper part of this figure, the four fixed vacua $1$, $2$, $3$ and $4$, present for all values of $\sigma$, have been depicted in black, while the variable vacua are depicted in red. In sum, the vacuum manifold is composed of a different number of vacua depending on the value of $\sigma$:
\begin{itemize}
	\item For values of the parameter $\sigma<\frac{3}{2}$, $3$ variable vacua appear and the vacuum manifold contains a total of $7$ vacua.
	\item In the case with $\sigma= \frac{3}{2}$, $3$ variable discrete vacua and a continuum of vacua at $\psi=- 1$ appears. Although the physical implications of the emergence of a continuous vacuum manifold is unclear, this case marks a clear transition between two distinct regimes.  
	\item When $\sigma>\frac{3}{2}$ new variable vacua appear, raising the total number to $13$.
\end{itemize} 

\begin{figure}[h!]
	\centering
	\includegraphics[height=4.5cm]{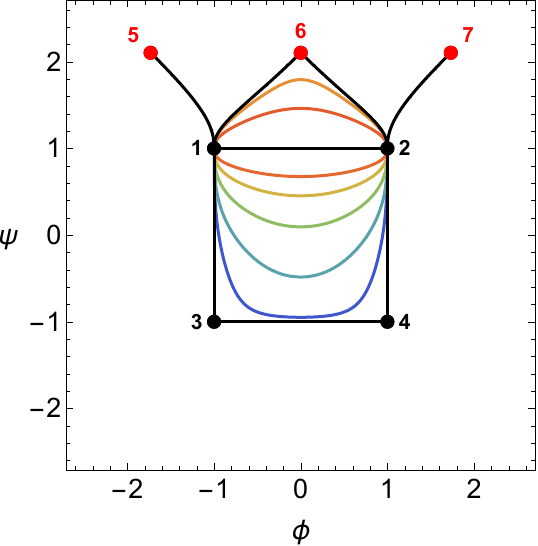}	\hspace{0.3cm}
	\includegraphics[height=4.5cm]{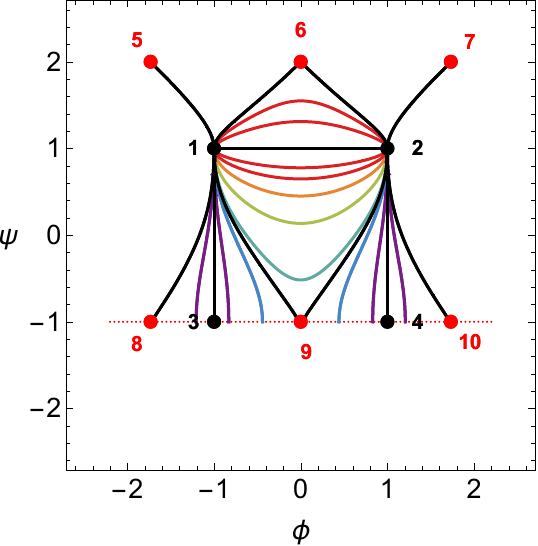} \hspace{0.3cm}
	\includegraphics[height=4.5cm]{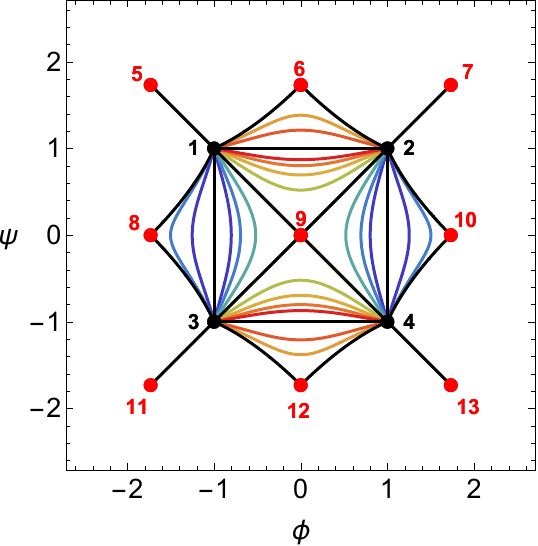}\\
	\includegraphics[height=4.5cm]{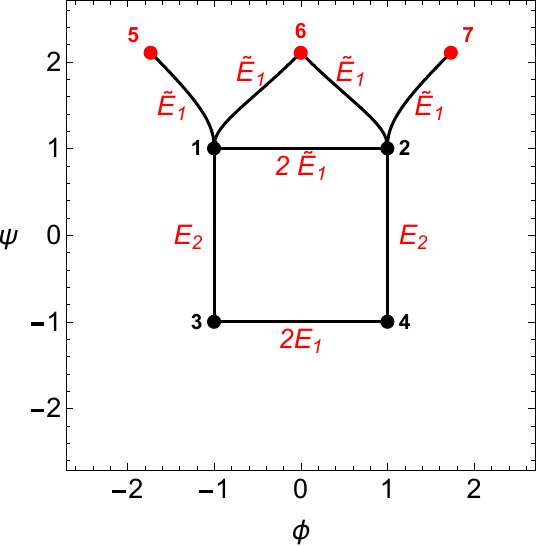}	\hspace{0.3cm}
	\includegraphics[height=4.5cm]{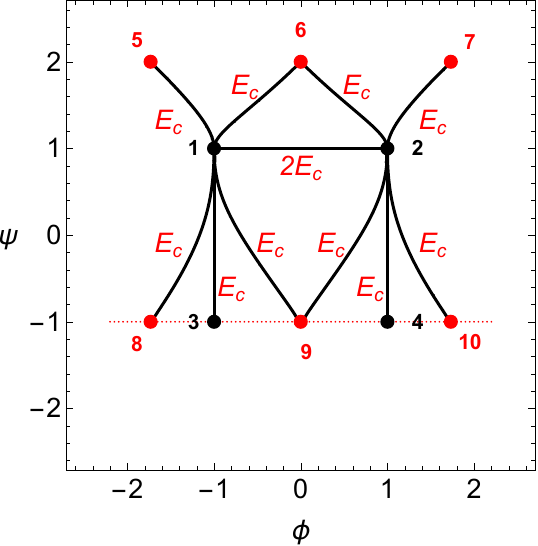} \hspace{0.3cm}
	\includegraphics[height=4.5cm]{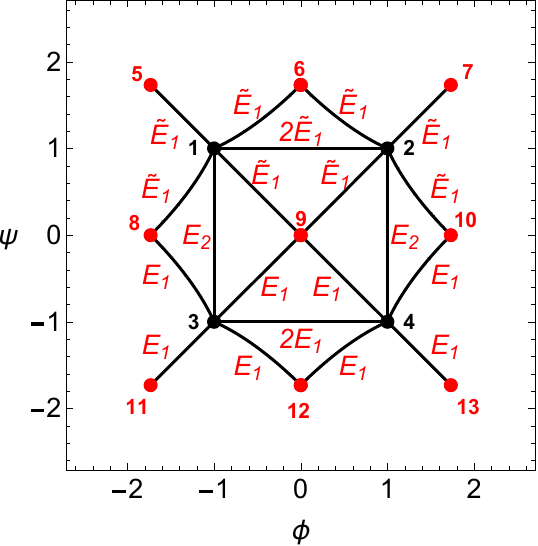}\\		
	\caption{\small Contourplot of the orbits of the emerging families of kinks (top panels) for different values of the constant $C$ for $\sigma=1$ (left panels), $\sigma=\frac{3}{2}$ (middle panels) and $\sigma=1000$ (right panels) and energies \eqref{eq:EnergiesFirst} of the boundary and isolated kinks (bottom panels). The four fixed vacua are represented in black while the variable ones are depicted in red. Boundary and isolated kinks are represented in black.}
	\label{fig:CompositePhi4Bvac}
\end{figure}

\noindent Now, even though directly obtaining solutions of the Euler-Lagrange equations \eqref{eq:feSigmamodel21}
\begin{align*}
	&-\frac{\partial^2\phi}{\partial t^2}+\frac{\partial^2\phi}{\partial x^2}=\phi (1 - \phi^2) \left( \sigma^2 \left(1 - \frac{\phi^2}{3}\right) (\psi^2 - 1)^2 - 
	2  \left(1 + \sigma \left(\psi - \frac{\psi^3}{3}\right)\right)^2 \right)\,,\\
	&-\frac{\partial^2\psi}{\partial t^2}+\frac{\partial^2\psi}{\partial x^2}=\sigma (1 - \psi^2) \left( (\phi^2 - 1)^2 \left(1 + \sigma \left(\psi - \frac{\psi^3}{3}\right) \right) -2 \sigma \left(\phi - \frac{\phi^3}{3}\right)^2 \psi \right)\,,
\end{align*}
may be challenging for non-vanishing $\sigma$, this family of potentials admits a family of superpotentials of the form
\begin{equation}
	W(\phi,\psi;\sigma)= \phi-\frac{\phi^3}{3} + \sigma \, \left(\phi-\frac{\phi^3}{3}\right)\left(\psi-\frac{\psi^3}{3}\right)\,.\label{eq:WPrimerModelo}
\end{equation}
Notice that this superpotential vanishes at the variable vacua, defined by equation \eqref{eq:cubic}. Therefore, BPS kinks that asymptotically connect a variable vacuum point must also connect a fixed vacuum point at the other end. This also implies that the energy \eqref{eq:EnergyBPS1} of such BPS kinks will be determined by the value of the superpotential at the fixed vacua. Now, static kinks can be identified by solving the pair of Bogomol'nyi equations \eqref{bpsb1} associated with the superpotentials \eqref{eq:WPrimerModelo}
\begin{equation}\label{eq:Bogomolnyiphi4b}
	\frac{d\phi}{dx}= \pm \left[1+\sigma\left(\psi-\frac{\psi^3}{3}\right)\right] ~ (\phi^2-1) \, , \qquad \frac{d\psi}{dx}= \pm \left[\sigma \left(\phi-\frac{\phi^3}{3}\right)\right] ~ (\psi^2-1) \, .
\end{equation}
It should be noted that, while these equations share terms with the Bogomol'nyi equations of a double $\phi^4-$model, the solutions will differ significantly as new zeros appear on the right-hand side. On one hand, kinks can be sought by imposing the trial orbits $\phi=\pm 1$ or $\psi=\pm 1$. Under a suitable reparametrization of the spatial axis $x$, these conditions lead to solutions of the $\phi^4-$model in both fields. For the component $\phi$ this reparametrization is possible as long as $\sigma\neq\frac{3}{2}$, while for the component $\psi$ it is possible only when $\sigma\neq 0$. These and other special kinks shall be denoted according to the vacua they connect. That is, a kink connecting vacua $i$ and $j$ will be denoted as $\Phi_{i,j}$. For instance, $\Phi_{1,2}$, given by $\psi=1$, will be of the form:
\begin{equation*}
	\Phi_{1,2}(x)=\left(\pm \tanh\left[\left(1-\frac{2}{3} \sigma\right) \left(x-x_0\right)\right],1\right)\,,
\end{equation*}
 with center of the kink $x_0 \in\mathbb{R}$. Moreover, the orbit flow equation derived from the Bogomol'nyi equations \eqref{eq:Bogomolnyiphi4b} can be solved
\begin{equation}\label{eq:ch6OrbitCompositeDoublephi4vac}
	\frac{d\psi}{d \phi}=\sigma \, \frac{\phi-\frac{\phi^3}{3}}{1+\sigma\left(\psi-\frac{\psi^3}{3}\right)} \, \frac{\psi^2-1}{\phi^2-1}\qquad \Rightarrow \qquad \left|\frac{1+\psi}{1-\psi}\right|^{\frac{3}{\sigma}-2} \left|1-\phi^2\right|^{\frac{1}{3}}=C \, e^{\frac{1}{6}\left(\phi^2-\psi^2\right)}\, ,
\end{equation}
where $C>0$ is a constant that distinguishes between members of the arising family of kinks for the model with value $\sigma$. Let us denote this one-parameter family of kinks as $\Sigma_{i,j}(C)$. Depending on the value of $\sigma$ different regimes of solutions arise, in which the symmetry $\phi\rightarrow -\phi$ exhibited by the potential \eqref{eq:PrimerPotencial} is clearly visible:
\begin{itemize}
	\item When $0<\sigma<\frac{3}{2}$ four kinks $\Phi_{1,2}$, $\Phi_{2,4}$, $\Phi_{3,4}$ and $\Phi_{1,3}$ appear forming the edges of a square with vertices at points $1$, $2$, $3$ and $4$, which shall be denoted as $S$. Moreover, outside this square $S$, two ``isolated'' kinks, $\Phi_{1,5}$ and $\Phi_{2,7}$, as well as two ``boundary'' kinks, $\Phi_{1,6}$ and $\Phi_{2,6}$, emerge. These two boundary kinks, together with $\Phi_{1,3}$, $\Phi_{3,4}$ and $\Phi_{2,4}$, form a region densely filled with a whole family of kinks, $\Sigma_{1,2}(C)$, which connects the vacua $1$ and $2$, see Figure \ref{fig:CompositePhi4Bvac} (left). Different colors have been employed to label solutions with different integration constant $C$. Inside $S$, limits of the constant $C\rightarrow\infty$ and $C\rightarrow 0$ correspond to the kinks $\Phi_{1,3}$, $\Phi_{3,4}$ and $\Phi_{2,4}$ and to the kink $\Phi_{1,2}$ respectively. Outside $S$, $C$ can only reach a maximum value that corresponds to the boundary kinks $\Phi_{1,6}$ and $\Phi_{2,6}$. 
	\item When $\sigma=\frac{3}{2}$ the structure of the kink variety changes significantly due to the appearance of the three variable vacua $8$, $9$ and $10$, as well as the continuum of vacua along $\psi=-1$. On one hand, this continuum of vacua causes the boundary kink from the previous case $\Phi_{3,4}$ to disappear, see Figure \ref{fig:CompositePhi4Bvac} (middle). On the other, the family of kinks behaves differently in three distinct regions. The region enclosed by boundary kinks $\Phi_{1,6}$, $\Phi_{1,9}$, $\Phi_{2,6}$ and $\Phi_{2,9}$ is densely filled by a family of kinks $\Sigma_{1,2}(C)$, which remains in the same topological sector as in the case $0<\sigma<\frac{3}{2}$. These kinks are shown once again with a color that depends on the value of $C$. For solutions with $\psi>1$ there is a maximum value of $C$, which corresponds to boundary kinks $\Phi_{1,6}$ and $\Phi_{2,6}$. However, in the regions between $\Phi_{1,8}$, $\Phi_{1,9}$ or $\Phi_{2,9}$, $\Phi_{2,10}$ and the continuum of vacua, vacuum $1$ or $2$ is asymptotically connected to different points of that continuum. It is worth noting that the isolated kinks $\Phi_{1,5}$ and $\Phi_{2,7}$ remain present throughout this ``transition''.
	
	\item Lastly, for $\sigma>\frac{3}{2}$ new vacua emerge, which lead to the formation of new several boundary and isolated kinks. In particular, the vacuum point $9$ appears inside the square $S$ forming boundary kinks $\Phi_{1,9}$, $\Phi_{2,9}$, $\Phi_{3,9}$ and $\Phi_{4,9}$ that cross the square, splitting $S$ into four regions, see Figure \ref{fig:CompositePhi4Bvac} (right). Consequently, four different families of kinks $\Sigma_{1,2}(C)$, $\Sigma_{2,4}(C)$, $\Sigma_{3,4}(C)$ and $\Sigma_{1,3}(C)$ in different topological sectors appear densely filling these regions. This also occurs outside the square, where four densely filled ``triangles'' appear in the regions between $S$ and the boundary kinks $\Phi_{1,6}$, $\Phi_{1,8}$, $\Phi_{2,6}$, $\Phi_{2,10}$, $\Phi_{3,8}$, $\Phi_{3,12}$, $\Phi_{4,10}$, $\Phi_{4,12}$. It is also worth highlighting that the boundary kinks lost in the ``transition'' are recovered, since the reparametrization is once more possible. Moreover, instead of two, four isolated kinks $\Phi_{1,5}$, $\Phi_{2,7}$, $\Phi_{3,11}$ and $\Phi_{4,13}$ emerge in this scenario. Note that this distribution of kinks respects the underlying symmetry $\phi\rightarrow-\phi$, which relates boundary kinks, isolated kinks, as well as families of kinks, with positive and negative $\phi$.
\end{itemize}

On the other hand, the energy of the boundary and isolated kinks in the three scenarios can be easily computed \eqref{eq:EnergyBPS1}, see Figure \ref{fig:CompositePhi4Bvac} for their distribution, where labels correspond to energies 
\begin{equation}\label{eq:EnergiesFirst}
	E_1=\left|\frac{2}{3}-\frac{4 \sigma}{9}\right|\,, \qquad \widetilde{E}_1=\left|\frac{2}{3}+\frac{4 \sigma}{9}\right|\, \qquad E_2=\left|\frac{8 \sigma}{9}\right|\,, \qquad E_c=\frac{4}{3}\,.
\end{equation}
Since the superpotential vanishes at all variable vacua, the energy of all kinks that joins one variable vacuum point with one vacuum point in $S$ is determined by the vacua in $S$ to which it is asymptotically connected. Moreover, it is worth noting the emergence of the relation between energies $\widetilde{E}_1=E_1+E_2$ for $0<\sigma<\frac{3}{2}$ and the relation $E_2=\widetilde{E}_1+E_1$ when $\sigma>\frac{3}{2}$. Let us consider the three scenarios described above separately:
\begin{itemize}
	\item When $0<\sigma<\frac{3}{2}$  the energies of boundary and isolated kinks are given by:
	\begin{align*}
		& E\left[\Phi_{1,2}\right]= 2 \widetilde{E}_1\,, \quad E\left[\Phi_{1,3}\right]=E\left[\Phi_{2,4}\right]= 2 E_2\,, \quad E\left[\Phi_{3,4}\right]=2 E_1\,,\\
		& E\left[\Phi_{1,5}\right]=E\left[\Phi_{1,6}\right]=E\left[\Phi_{2,6}\right]=E\left[\Phi_{2,7}\right]=\widetilde{E}_1\,.
	\end{align*}
	 The energy of members of the family of kinks $\Sigma_{1,2}(C)$ is also that of the boundary kink $\Phi_{1,2}$ with which shares topological sector, which depends on $\sigma$:
	\begin{equation*}
		E\left[\Sigma_{1,2}\right]=E\left[\Phi_{1,2}\right]=E\left[\Phi_{1,6}\right]+E\left[\Phi_{2,6}\right]=E\left[\Phi_{1,3}\right]+E\left[\Phi_{2,4}\right]+E\left[\Phi_{3,4}\right]=2\widetilde{E}_1\,,
	\end{equation*}
	 where the energy sum rules that arise in this regime are written explicitly. It is worth highlighting that for $\sigma>0$ the energy $E\left[\Sigma_{1,2}\right]$ is larger than that of the $\phi^4-$model.
	\item When $\sigma=\frac{3}{2}$ the energy of the boundary kink $\Phi_{1,2}$ is twice of that of the $\phi^4-$model $E\left[\Phi_{1,2}\right]=2 E_c=\frac{8}{3}\,$, while the rest of the boundary and isolated kinks share the same energy $E_c=\frac{4}{3}\,$, which is that of the kinks of the $\phi^4-$model. Accordingly, the energy of the members of the family $\Sigma_{1,2}(C)$ is also twice of that of the $\phi^4-$model:
	\begin{equation*}
		E\left[\Sigma_{1,2}\right]=E\left[\Phi_{1,2}\right]=E\left[\Phi_{1,6}\right]+E\left[\Phi_{2,6}\right]=E\left[\Phi_{1,9}\right]+E\left[\Phi_{2,9}\right]=2 E_c\,.
	\end{equation*}
	Indeed, $\sigma=\frac{3}{2}$ implies that $E_1=0$ and $\widetilde{E}_1=E_2=E_c$. The energy of the rest of the kinks that join vacua $1$ or $2$ to the continuum of vacua is also $E_c$. 
	\item When $\sigma>\frac{3}{2}$ the energy of the boundary kinks that form $S$ are given by
	\begin{align*}
		E\left[\Phi_{1,2}\right]= 2 \widetilde{E}_1\,, \qquad E\left[\Phi_{1,3}\right]=E\left[\Phi_{2,4}\right]= 2 E_2 \,, \qquad E\left[\Phi_{3,4}\right]= 2 E_1\,.
	\end{align*}
	Two different values of energy, $E_1$ and $\widetilde{E}_1$, appear for the rest of boundary and isolated kinks
	\begin{align*}
		&E\left[\Phi_{1,5}\right]=E\left[\Phi_{1,6}\right]=E\left[\Phi_{1,8}\right]=E\left[\Phi_{1,9}\right]=E\left[\Phi_{2,6}\right]=E\left[\Phi_{2,7}\right]=E\left[\Phi_{2,9}\right]=E\left[\Phi_{2,10}\right]=\widetilde{E}_1\,,\\
		&E\left[\Phi_{3,8}\right]=E\left[\Phi_{3,9}\right]=E\left[\Phi_{3,11}\right]=E\left[\Phi_{3,12}\right]=E\left[\Phi_{4,9}\right]=E\left[\Phi_{4,10}\right]=E\left[\Phi_{4,12}\right]=E\left[\Phi_{4,13}\right]=E_1\,.
	\end{align*}
	 Lastly, the four emerging families of kinks present energies and energy sum rules:
	 \begin{align*}
	 	&E\left[\Sigma_{1,2}\right]=E[\Phi_{1,2}]=E[\Phi_{1,6}]+E[\Phi_{2,6}]=E[\Phi_{1,9}]+E[\Phi_{2,9}]=2 \widetilde{E}_1\,,\\
	 	&E\left[\Sigma_{1,3}\right]=E[\Phi_{1,3}]=E[\Phi_{1,8}]+E[\Phi_{3,8}]=E[\Phi_{1,9}]+E[\Phi_{3,9}]=E_2\,,\\
	 	&E\left[\Sigma_{2,4}\right]=E[\Phi_{2,4}]=E[\Phi_{2,10}]+E[\Phi_{4,10}]=E[\Phi_{2,9}]+E[\Phi_{4,9}]=E_2\,,\\
	 	&E\left[\Sigma_{3,4}\right]=E[\Phi_{3,4}]=E[\Phi_{3,12}]+E[\Phi_{4,12}]=E[\Phi_{3,9}]+E[\Phi_{4,9}]=2 E_1\,,
	 \end{align*}
\end{itemize}

It is also worth noting that, apart from the symmetry $\phi\rightarrow-\phi$, another symmetry $(\psi,-\sigma)\rightarrow(-\psi,\sigma)$ appears in Bogomol'nyi equations \eqref{eq:Bogomolnyiphi4b}. Indeed, cases with negative values of $\sigma$ can be obtained by inverting vertically the position of vacua and kinks,. This also explains the symmetrical distribution of energies, since the symmetry $(\psi,-\sigma)\rightarrow(-\psi,\sigma)$ transforms $E_1\leftrightarrow \widetilde{E}_1$ and $E_2\leftrightarrow E_2$ as expected.

The fact that kinks in this model can be analytically identified is due to the fact that a superpotential \eqref{eq:WPrimerModelo} can be found. Moreover, the form of these potentials reveals a procedure for shaping the vacuum manifold and imposing constraints on the kink variety.  On one hand, the arising vacuum manifold inherits the usual vacua $1$, $2$, $3$ and $4$ present in the double $\phi^4-$model. On the other, after reparametrizations in the spatial axis, solutions of the $\phi^4-$model are retrieved as boundary kinks joining these vacua. This is due to the fact that this superpotential is written as a combination of superpotentials of $\phi^4-$models. This allows to inherit certain structure in the vacuum manifold and in turn in the kink variety. However, unlike the double $\phi^4-$model, the kink variety of these models is not confined in the central square $\phi,\psi\in(-1,1)$. Furthermore, even in the central area the behavior of solutions does not resemble that of the $\phi^4-$model \cite{Manton2004,Shnir2018}, see orbits of solutions depicted in Figure \ref{fig:CompositePhi4Bvac}. This discrepancy arises because the model is a deformation of a double $\phi^4-$model. Indeed, the potential \eqref{eq:PrimerPotencial} can be written as
\begin{align*}
	&V(\phi,\psi;\sigma)=\frac{1}{2}\left(1-\phi^2\right)^2+ \sigma \, \left[ \left(\psi-\frac{\psi^3}{3}\right)\left(1-\phi^2\right)^2\right]+\\
	&+ \sigma^2 \, \left[ \left(\psi-\frac{\psi^3}{3}\right)^2\frac{1}{2}\left(1-\phi^2\right)^2+ \left(\phi-\frac{\phi^3}{3}\right)^2\frac{1}{2}\left(1-\psi^2\right)^2\right]	\,,
\end{align*}
where $\sigma$ takes the role of the deformation parameter. Lastly, the intricacies of this model motivates us to generalize this procedure in the next section, enabling new types of composition of field theories.

\section{Composite scalar field theories}

In the previous section, the properties emerging from the analyzed model suggested that a more general formalism could be developed for generating field theories that inherit certain analytical solutions. This is the aim of the present section, where a generalized method will be introduced. In particular, given two field theories with target spaces $\mathbb{R}^{m}$ and $\mathbb{R}^{n}$, we shall construct a new field theory with target space $\mathbb{R}^{m+n}$, in which the corresponding fields are coupled, extending the dynamics of the two originally independent target subspaces $\mathbb{R}^{m}$ and $\mathbb{R}^{n}$. This procedure will be performed to entangle of their dynamics while ensuring that certain solutions are inherited in the resulting field theory. 

Given a configuration $\Omega:\mathbb{R}^{1,1}\rightarrow \mathbb{R}^{m+n}$ for the new field theory, let us decompose its action in coordinates into the kinetic parts of the original field theories on $\mathbb{R}^m$ and $\mathbb{R}^n$ and a potential function $V$ that intertwines the dynamics of the two subspaces $\mathbb{R}^m$ and $\mathbb{R}^n$ by coupling their fields  
$$S\left[\Omega\right]=\int_{\mathbb{R}^{1,1}}\left[\frac{1}{2} ~  \eta^{\alpha \beta}\sum_{i=1}^{m}\frac{d\phi^i}{dx^{\alpha}}\frac{d\phi^i}{dx^{\beta}}+\frac{1}{2} ~  \eta^{\alpha \beta}\sum_{j=1}^{n}\frac{d\psi^j}{dx^{\alpha}}\frac{d\psi^j}{dx^{\beta}}-V(\phi, \psi)\right] ~ dx dt \, ,$$
where Einstein convention has been used for greek indices,  $\eta={\rm Diag}(1,-1)$, Cartesian coordinates in the internal space have been denoted as $\left\{\phi,\psi\right\}=\{\phi^1,\dots,\phi^m,\psi^1,\dots,\psi^n\}$ and where $\{\phi^i\}$ with $i=1,\dots,m$ are the coordinates for $\mathbb{R}^m$ and $\{\psi^j\}$ with $j=1,\dots,n$ those for $\mathbb{R}^n$. This intertwinedness in the action is inherited by the field equations, which can also be split in the two sets
\begin{align}
	& -\frac{\partial^2\phi^i}{\partial t^2}+\frac{\partial^2\phi^i}{\partial x^2}= \frac{\partial V}{\partial \phi^i} \, , \quad i=1, \dots, m \, ,\label{eq:feSigmamodel1b}\\
	&-\frac{\partial^2\psi^i}{\partial t^2}+\frac{\partial^2\psi^i}{\partial x^2}=\frac{\partial V}{\partial \psi^j}\, , \,  \, j=1,\dots, n \, . \label{eq:feSigmamodel2b}
\end{align}
 Notice that if the potential can be decomposed as $V(\phi,\psi)=V_A(\phi)+V_B(\psi)$, then equations are decoupled and the two independent original field theories are recovered. While in the decoupled scenario the vacuum manifold is simply the Cartesian product of the original ones $\mathcal{M}=\mathcal{M}_A \times \mathcal{M}_B$, a more general potential gives rise to a more intricate vacuum manifold structure.
 
The energy for a given configuration $\Omega:\mathbb{R}^{1,1}\rightarrow \mathbb{R}^{m+n}$ can be written as
\begin{equation*}
	E[\phi]= \int^{\infty}_{-\infty}  \, \left[ \frac{1}{2}\, \sum_{i=1}^{m}\left[\left(\frac{\partial\phi^i}{\partial t}\right)^2+\left(\frac{\partial\phi^i}{\partial x}\right)^2\right]+ \frac{1}{2}\, \sum_{j=1}^{n}\left[\left(\frac{\partial\psi^j}{\partial t}\right)^2+\left(\frac{\partial\psi^j}{\partial x}\right)^2\right] + V(\phi,\psi)\right] \, dx \,.
\end{equation*}
 Let us restrict to potentials $V(\phi,\psi)$ that can be written in terms of a superpotential $W(\phi,\psi):\mathbb{R}^{m+n}\rightarrow\mathbb{R}$ as follows
\begin{equation}
	V(\phi,\psi)= \frac{1}{2}\sum_{i=1}^{m} \left(\frac{\partial W}{\partial \phi^i} \right)^2 + \frac{1}{2}\sum_{j=1}^{n} \left(\frac{\partial W}{\partial \psi^j} \right)^2\, ,  \label{ps}
\end{equation}
with $i=1,\dots,m$  and $j=1,\dots,n$, so that the Bogomol'nyi arrangement can be performed and the following pair of sets of Bogomol'nyi equations can be obtained:
\begin{equation}
	\frac{d\phi^i}{dx}=(-1)^{\epsilon} \,  \frac{\partial W}{\partial \phi^i} \, ,\qquad \frac{d\psi^j}{dx}=(-1)^{\epsilon}  \, \frac{\partial W}{\partial \psi^j} \, . \label{bpsb}
\end{equation}
 Once again, solutions of these Bogomol'nyi equations will be solutions of \eqref{eq:feSigmamodel1b} and \eqref{eq:feSigmamodel2b}, and will lead to the topological charge
\begin{equation}\label{eq:EnergyBPSb}
	T=\Big\vert \lim_{x\rightarrow \infty} W[\phi(x),\psi(x)]- \lim_{x\rightarrow -\infty}W[\phi(x),\psi(x)]\Big\vert \, ,
\end{equation}
which corresponds to the static energy of the kink.

  Let us now consider two real scalar field theories with respective fields $\phi^1,\dots,\phi^m$ and $\psi^1,\dots,\psi^m$, potentials $V_A(\phi)$ and $V_B(\psi)$ and  actions
  \begin{equation*}
  	S_A\left[\phi\right]=\int_{\mathbb{R}^{1,1}}\left[\frac{1}{2} ~  \eta^{\alpha \beta}\sum_{i=1}^{m}\frac{d\phi^i}{dx^{\alpha}}\frac{d\phi^i}{dx^{\beta}}-V_A(\phi)\right]dx dt \,, \, S_B\left[\psi\right]=\int_{\mathbb{R}^{1,1}}\left[\frac{1}{2} ~  \eta^{\alpha \beta}\sum_{j=1}^{n}\frac{d\psi^j}{dx^{\alpha}}\frac{d\psi^j}{dx^{\beta}}-V_B(\psi)\right]dx dt.
  \end{equation*} 
  Moreover, let us restrict to field theories that admit superpotentials $W_A(\phi)$ and $W_B(\psi)$ for $V_A(\phi)$ and $V_B(\psi)$ respectively:
  \begin{equation*}
  	V_A(\phi)=\frac{1}{2}\sum_{i=1}^{m} \left(\frac{\partial W_A}{\partial \phi^i} \right)^2\,, \qquad V_B(\psi)= \frac{1}{2}\sum_{j=1}^{n} \left(\frac{\partial W_B}{\partial \psi^j} \right)^2\, ,
  \end{equation*} 
  In order to be able to impose certain conditions on the kink variety, the superpotential of the new field theory $W(\phi,\psi)$ will be constructed from these two other superpotentials $W_A(\phi)$ and $W_B(\psi)$. This is, given two superpotentials $W_A:\mathbb{R}^m\rightarrow\mathbb{R}$ and $W_B:\mathbb{R}^n\rightarrow\mathbb{R}$ of two field theories, then a family of superpotentials will be defined on the product target space $W:\mathbb{R}^{m+n}\rightarrow \mathbb{R}$	as
\begin{equation}\label{eq:ExtensionSuper}
	W(\phi,\psi)=\alpha \, W_A(\phi)+\beta \,W_B(\psi)+ \sigma ~ W_A(\phi) ~ W_B(\psi)\, ,
\end{equation}
where $\alpha,\beta,\sigma\in \mathbb{R}$. This family of superpotentials \eqref{eq:ExtensionSuper} defines a family of potentials on the product space $V:\mathbb{R}^{m+n}\rightarrow \mathbb{R}$
\begin{equation}\label{eq:ch6potentialComposite}
	V(\phi,\psi;\alpha,\beta,\sigma)= (\alpha+\sigma \, W_B(\psi))^2 ~ V_A(\phi)+ (\beta+\sigma \, W_A(\phi))^2 ~ V_B(\psi)\, ,
\end{equation}
which not only inherits a combination of the original vacua from the field theories defined on the subspaces, but also potentially engenders new ones. It is worth highlighting that in the model presented in Section $2$, the inherited and extra vacua were referred to as fixed and variable vacua, respectively. Notice that continuous vacua can appear when at least one of the following conditions hold
\begin{align}
	& W'_A(\phi)=\alpha +\sigma \, W_A(\phi)=0,\label{eq:Degenerate1}\\
	& W'_B(\psi)=\beta +\sigma \, W_B(\psi)=0.\label{eq:Degenerate2}
\end{align}
 This is, when the components of the extra vacua in the subspaces $\mathbb{R}^n$ or $\mathbb{R}^m$ are vacua of the original models, $V_A(\phi_{v,extra})= 0$ and/or $V_B(\psi_{v,extra})= 0$, continuous vacua appear. Notice that potential \eqref{eq:ch6potentialComposite} can be seen as a second order expansion on the parameter $\sigma$ around the uncoupled case:
\begin{align}
	V(\phi,\psi;\alpha,\beta,\sigma)&=\alpha \,V_A(\phi)+\beta \,V_B(\psi)+ \sigma \, \left(2 \alpha W_B(\psi)V_A(\phi)+2 \beta W_A(\phi)V_B(\psi)\right)+\notag\\
	&+ \sigma^2 \, \left( W_B^2(\psi)V_A(\phi)+ W_A^2(\phi)V_
	B(\psi)\right)\,.	
\end{align}

 On the other hand, the Hessian of the superpotential carries information about the stability of solutions of the gradient system \eqref{bpsb}. Since the new superpotential \eqref{eq:ExtensionSuper} is written in terms of the original ones, $W_A$ and $W_B$, the eigenvalues of the Hessian at the inherited and extra vacua are fixed. Let us denote as $\lambda_i$ and $\lambda_j$ with $i=1,\dots,n$ and $j=1,\dots,m$ the eigenvalues of the Hessian in the respective subspaces $\mathbb{R}^n$ and $\mathbb{R}^m$.  On one hand, it is easy to check that at an inherited vacuum point, $(\phi_{v,inh},\psi_{v,inh})$, the eigenvalues of the original models $\lambda_{i,A}$ and $\lambda_{j,B}$ are recovered up to two constants
 \begin{equation}\label{eq:EigenvaluesInherited}
 	\lambda_i=(\alpha +\sigma \, W_B(\psi_{v,inh})) \, \lambda_{i,A} \,, \qquad \lambda_j=(\beta +\sigma \, W_A(\phi_{v,inh})) \, \lambda_{i,B} \,, 
 \end{equation}
 which may change the stability of solutions in the new model. Notice that these constants vanish at points where continuous vacua emerge \eqref{eq:Degenerate1} and \eqref{eq:Degenerate2}, leading to degenerate critical points of the superpotential. Furthermore, saddle points in the original models, if any, are inherited by the new model if the preceding constants do not vanish. It is worth highlighting that different combinations of eigenvalues of the original models will give rise to inherited vacua that are minima, maxima and saddle points in the combined model. However, for the extra vacua, two non-zero eigenvalues of opposite signs are always identified
 \begin{equation*}
 	\lambda=\pm 2\sigma \, \sqrt{V_A(\phi_{v,extra}) V_B(\psi_{v,extra})}\,,
 \end{equation*} 
 as well as $m+n-2$ zero eigenvalues. In particular, when $m=n=1$ extra points are saddle points of the superpotential \eqref{eq:ExtensionSuper}. It is also worth noting that when all critical points are non-degenerate and all intersections of stable and unstable manifolds are transverse, the presence of $n-$parameter families of kinks can be excluded between certain critical points. This will be analyzed in the next section.

  A straightforward calculation shows that if extra vacua $v_{extra}$ appear, then the superpotential \eqref{eq:ExtensionSuper} must have the same value at those points
  \begin{equation}\label{eq:SuperpotentialExtraVacua}
  	W(v_{extra})=-\frac{\alpha \beta}{\sigma}\,.
  \end{equation} 
  This is also true for any vacua in the continua given by \eqref{eq:Degenerate1} and \eqref{eq:Degenerate2}. This implies, as expected, given that the energy of BPS kinks is determined by \eqref{eq:EnergyBPSb}, that BPS kinks must connect at least one inherited vacuum. On the other, it follows that all emerging kinks asymptotically connecting a given inherited vacuum to any extra vacuum point have the same energy. Moreover, hyperplanes $\phi^i=\phi_{v,i}$ and $\psi^j=\psi_{v,i}$ split $\mathbb{R}^{m+n}$ in regions where at most one extra vacua can appear. Indeed, inherited vacua are critical points of the terms $\alpha+\sigma \, W_B$ and $\alpha+\sigma \, W_A$, whose zeros define the location of the extra vacua. These extra vacua must therefore be isolated. Hence, all the information about the energy of these kinks is contained in the values of the parameters fixed by the model $\alpha,\beta,\sigma$ and in the values of the superpotential at the inherited vacua. Let us examine these three scenarios:
\begin{itemize}
	\item BPS kinks between inherited vacua $(\phi_{v,1},\psi_{v,1})$ and $(\phi_{v,2},\psi_{v,2})$ have an energy:
	\begin{equation*}
		E=\left|\alpha \, \Delta W_A+\beta \, \Delta W_B+\sigma \left[W_A(\phi_{v,1}) W_B(\psi_{v,1})-W_A(\phi_{v,2}) W_B(\psi_{v,2})\right]\right|\,,
	\end{equation*}
	where $\Delta W_A=W_A(\phi_{v,1})-W_A(\phi_{v,2})$ and $\Delta W_B=W_B(\psi_{v,1})-W_B(\psi_{v,2})$. Notice that when a kink between two inherited vacua satisfies 
	\begin{equation}\label{eq:ConditionSameEnergy}
		W_A(\phi_{v,1}) W_B(\psi_{v,1})=W_A(\phi_{v,2}) W_B(\psi_{v,2})\,,
	\end{equation}
	the energy of these kinks is independent of $\sigma$, and therefore is shared by the whole family of models. Moreover, it also follows that if the fields associated with one the subspaces lie at a vacuum of an original model, then the energy of the corresponding kink is directly related to that of the original theory:
	\begin{equation}
		E_A=\left|\alpha + \sigma\, W_B(\psi_{v,i})\right| \, E_{A,0} \,, \qquad E_B=\left|\alpha + \sigma\, W_A(\phi_{v,j})\right| \, E_{B,0} \,,
	\end{equation}
	for $i$ and $j$ from $1$ to the number of vacua in the original models $N_A$ and $N_B$ respectively and where $E_{A,0}$ and $E_{B,0}$ are the energies of the respective original kinks. Therefore, these inherited kinks will present up to $N_A+N_B$ different energies.  
	\item In this context, BPS kinks between extra vacua are forbidden, because these vacua share the same value of the superpotential \eqref{eq:SuperpotentialExtraVacua} and no gradient flow between them is possible.
	\item BPS kinks between an inherited vacuum $(\phi_{v},\psi_{v})$ and an extra vacua have an energy 
	\begin{equation}
		E=\left|\alpha \, W_A(\phi_v)+\beta \, W_B(\psi_v)+\sigma \, W_A(\phi_{v}) W_B(\psi_{v})+ \frac{\alpha \beta}{\sigma}\right|\,,
	\end{equation}
	which depend on $\sigma$ unless $W_A(\phi_{v})$ and/or $W_B(\psi_{v})$ vanish and simultaneously $\alpha$ and/or $\beta$ vanish.
\end{itemize}

On the other hand, this family of superpotentials \eqref{eq:ExtensionSuper} gives rise to the following Bogomol'nyi equations
\begin{equation}\label{eq:ch6BogomolnyiComposite}
	\frac{d\phi^i}{dx}= \pm \left[\alpha+\sigma W_B\right] ~ \frac{\partial W_A}{\partial\phi^i}\, ,  \qquad \frac{d\psi^j}{dx}= \pm \left[\beta+\sigma W_A\right] ~ \frac{\partial W_B}{\partial\psi^j} \, .
\end{equation}
While the orbit flow equations restricted to subspaces $\mathbb{R}^m$ and $\mathbb{R}^n$ are identical to those of the original field theories
$$\frac{d\phi^i}{d\phi^j} =\frac{\frac{\partial W_A}{\partial\phi^i}}{\frac{\partial W_A}{\partial\phi^j}} \, , \qquad \frac{d\psi^i}{d\psi^j} =\frac{\frac{\partial W_B}{\partial\psi^i}}{\frac{\partial W_B}{\partial\psi^j}} \, , \qquad i\neq j \,,$$	
solving the orbit flow equations that mix coordinates on $\mathbb{R}^m$ and $\mathbb{R}^n$ will be challenging in general
\begin{equation}\label{eq:OrbitFlowEqsMixing}
	\frac{\beta+\sigma W_A}{\frac{\partial W_A}{\partial\phi^i}} d\phi^i= \frac{\alpha+\sigma W_B}{\frac{\partial W_B}{\partial\psi^j}} d\psi^j\, .
\end{equation}
Even in the coupled case $\sigma\neq 0$, vacuum solutions of one of the original field theory make one of the equations \eqref{eq:ch6BogomolnyiComposite} trivially hold, while the other equation corresponds to that of the other original model up to a reparametrization by a constant. Indeed, if we consider a vacuum solution $\phi_v\in \mathcal{M}_A$ in $\mathbb{R}^m$, then the first set of Bogomol'nyi equations hold and the second one becomes the original Bogomolnyi equations for $V_B$	
$$\frac{d\psi^j}{d\tilde{x}}= \pm~ \frac{\partial W_B}{\partial\psi^j}\,, $$
where a reparametrization by a constant has been performed:
\begin{equation}\label{eq:ReparametrisationGeneral}
	\tilde{x}= \left[\beta+\sigma W_A(\phi_v)\right] x \,.
\end{equation}
 It is also worth highlighting that for $\sigma\neq0$ the orbit flow equations \eqref{eq:OrbitFlowEqsMixing} that couple coordinates of both subspaces $\mathbb{R}^m$ and $\mathbb{R}^n$ depend only on the two parameter quotients $\frac{\alpha}{\sigma}$ and $\frac{\beta}{\sigma}$. 
 Depending on which parameters $\alpha,\beta,\sigma$ vanish, different interesting scenarios arise. In particular, we may consider the following two cases:
\begin{itemize}
	
	\item \textbf{Case with $\alpha=0$ or $\beta=0$}: The first scenario occurs when either $\alpha$ or $\beta$ vanishes. Since the case with $\alpha=0$ is completely analogue to that with $\beta=0$, let us focus only on the latter. When $\beta=0$, the model depends on two parameters
	\begin{align}\label{eq:Potentialbetacero}
		V(\phi,\psi;\alpha,\sigma)&=\alpha \,V_A(\phi)+ \sigma \, \left(2 \alpha W_B(\psi)V_A(\phi)\right)+ \sigma^2 \, \left( W_B^2(\psi)V_A(\phi)+ W_A^2(\phi)V_B(\psi)\right)	\,,
	\end{align}
	and when $\sigma=0$ the potential becomes independent of $\psi$. Although this case involves two parameters, the orbit flow equation that relates the coordinates of both subspaces
	\begin{equation}\label{eq:Orbitasbetacero}
		\frac{d\phi^i}{d\psi^j}= \frac{\frac{\alpha}{\sigma}+ W_B}{W_A} \frac{\frac{\partial W_A}{\partial\phi^i}}{\frac{\partial W_B}{\partial\psi^j}} \,,
	\end{equation} 
	shows that the behavior of the orbits depends only on the value of the quotient $\frac{\alpha}{\sigma}$. Notice that when $\alpha=0$ or $\beta=0$ the superpotential vanishes at the extra points \eqref{eq:SuperpotentialExtraVacua}. Moreover, $\alpha=\beta=0$ leads to a family of potential functions
	\begin{equation}
		V(\phi,\psi;\sigma)= \sigma^2 \, \left( W_B^2(\psi)V_A(\phi)+ W_A^2(\phi)V_B(\psi)\right)	\,,
	\end{equation}
	with $\sigma$ appearing as a global multiplicative factor that leaves the orbit of solutions unchanged. Indeed, the orbit flow equations are independent of this parameter.
	
	\item \textbf{Case with $\alpha=\beta=1$}: Another interesting scenario arises when $\alpha=\beta=1$, coupling the dynamics of two field theories with potentials $V_A(\phi)$ and $V_B(\psi)$ through the parameter $\sigma$:
	\begin{align}
		V(\phi,\psi;\sigma)&=V_A(\phi)+V_B(\psi)+ \sigma \, \left(2 W_B(\psi)V_A(\phi)+2  W_A(\phi)V_B(\psi)\right)+\notag\\
		&+ \sigma^2 \, \left( W_B^2(\psi)V_A(\phi)+ W_A^2(\phi)V_
		B(\psi)\right)\,.	
	\end{align}
	 In this scenario the potential is ``expanded'' around the uncoupled case. Note that the value of the superpotential \eqref{eq:SuperpotentialExtraVacua} at any extra vacua $W(v_{extra})=-\frac{1}{\sigma}$ decreases now as $|\sigma|$ increases.
	
\end{itemize}

This method of combining models, based on the extension of superpotentials, underlies the construction of the model presented in Section $2$. In that example, two $\phi^4-$models are combined with $\alpha=1$, $\beta=0$ and superpotentials:
\begin{equation}\label{eq:SuperpotentialsPhi4}
	W_A(\phi)=\phi-\frac{\phi^3}{3} \, , \qquad W_B(\psi)=\psi-\frac{\psi^3}{3}\, .
\end{equation}

 In the next sections, we explore different combinations of models. The kinks of the original field theories will be transferred to different ``composite'' models, shaping the remainder of the kink variety.

\subsection{Extension of the double $\phi^4-$model}

\begin{figure}[h!]
 	\centering
 	\includegraphics[height=4cm]{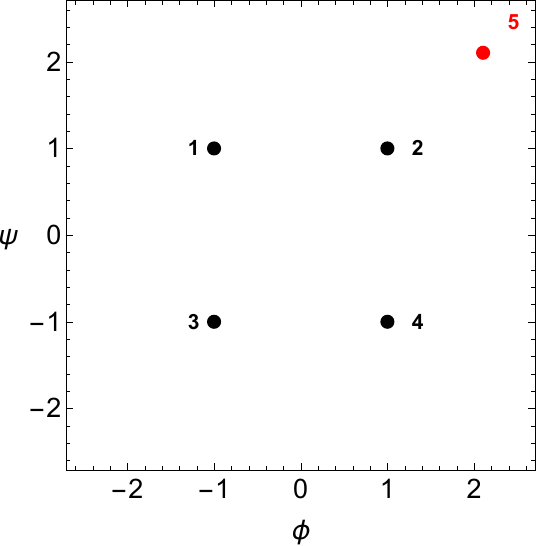}	\hspace{1.4cm}
 	\includegraphics[height=4cm]{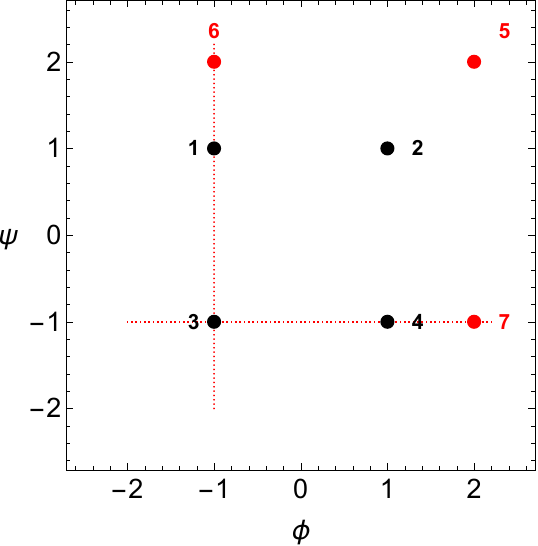} \hspace{1.4cm}
 	\includegraphics[height=4cm]{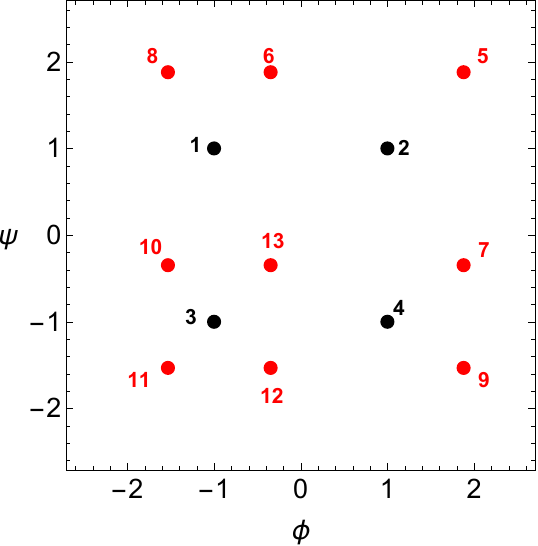}\\
 	\includegraphics[height=4cm]{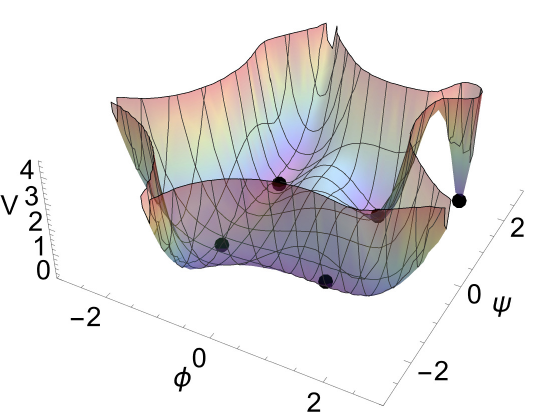}	\hspace{0.3cm}
 	\includegraphics[height=4cm]{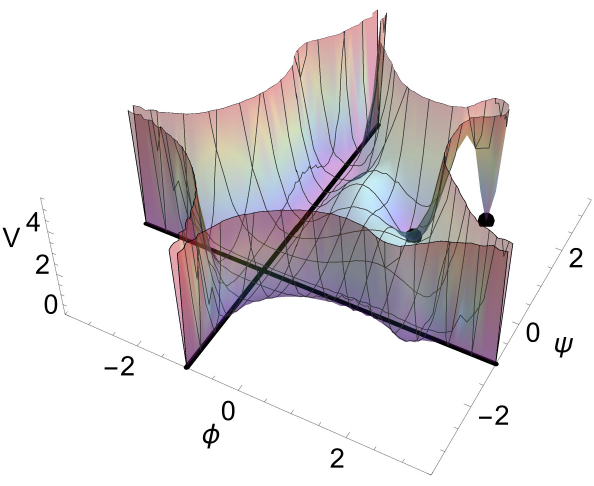} \hspace{0.3cm}
 	\includegraphics[height=4cm]{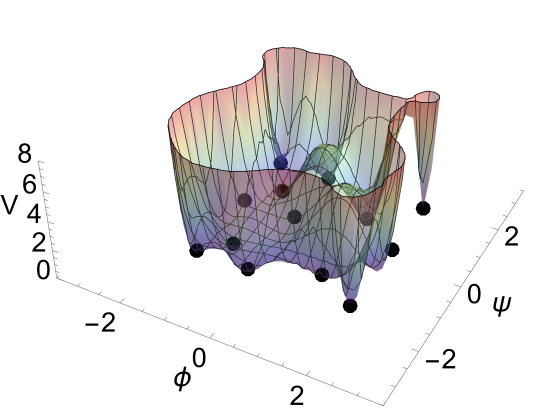}\\		
 	\caption{\small Vacua of the field theory (top panels) and the corresponding potential (bottom panels) for $\sigma=1$ (left), $\sigma=\frac{3}{2}$ (middle) and $\sigma=3$ (right). In the top panels, the four inherited vacua are shown in black while the extra ones are depicted in red.}
 	\label{fig:CompositePhi4bVacuaPotential}
 \end{figure}

As second example let us construct an extension of the double $\phi^4-$model using $\alpha=\beta=1$ and the same two superpotentials \eqref{eq:SuperpotentialsPhi4}. In this case the composite field theory on the plane $\mathbb{R}^2$ has a one-parameter family of superpotentials \eqref{eq:ExtensionSuper} given by
\begin{equation}\label{eq:SuperpotentialDoublePhi4}
	W=W_A(\phi) + W_B(\psi) + \sigma \, W_A(\phi) ~ W_B(\psi)\,,
\end{equation}
where once again the parameter $\sigma$ of the extension modulates the coupling. This superpotential, in turn, produces by equation \eqref{eq:ch6potentialComposite} a potential function $V:\mathbb{R}^2\rightarrow\mathbb{R}$ for each value of $\sigma$
\begin{equation}\label{eq:PotentialExtensionPhi4}
	V(\phi,\psi;\sigma)=\frac{1}{2} \left[1+\sigma\left(\psi-\frac{\psi^3}{3}\right)\right]^2  (1-\phi^2)^2 + \frac{1}{2} \left[1+\sigma \left(\phi-\frac{\phi^3}{3}\right)\right]^2  (1-\psi^2)^2 \, .
\end{equation} 
Given the relation $V(\phi,\psi;-\sigma)=V(-\phi,-\psi;\sigma)$, it suffices to consider non-negative $\sigma$, which contains all relevant information. Once more, combinations of the vacua of the original $\phi^4-$models generate four inherited vacua for all $\sigma$. However, the fact that $\beta \neq 0$ makes the vacuum manifold of this theory different from that of the first example. Now the extra vacua that emerge due to other factors in the potential \eqref{eq:PotentialExtensionPhi4} will be of the form $(p_i,p_j)$, where $p_i$ and $p_j$ are real solutions of the equation \eqref{eq:cubic}. As in the example shown in Section $2$, only one real solution of this equation is found when $0<\sigma<\frac{3}{2}$, two when $\sigma=\frac{3}{2}$ and three when $\sigma>\frac{3}{2}$. Consequently, the vacuum manifold once again contains a varying number of vacua, depending on the value of $\sigma$
\begin{equation*}
	\mathcal{M}_{\sigma}=\left\lbrace (-1,-1),(-1,1),(1,-1),(1,1),(p_i,p_j) \right\rbrace \, ,
\end{equation*}
 see Figure \ref{fig:CompositePhi4bVacuaPotential}. Once more, in the upper part of this figure the four inherited have been depicted in black, while the extra ones in red. Let us examine each case separately:
 \begin{itemize}
 	\item When $0<\sigma<\frac{3}{2}$ one extra vacuum point appears in the potential (point $5$) outside the square with vertices at points $1$, $2$, $3$ and $4$. Once again, this square shall be denoted as $S$. A total of $5$ vacua emerge in this scenario.
 	\item The case with $\sigma=\frac{3}{2}$ leads to one extra vacua (point $5$) outside the square $S$ and two combined continua of vacua that absorb inherited vacua $1$, $3$ and $4$. Indeed, both \eqref{eq:Degenerate1} and \eqref{eq:Degenerate2} vanish. In total, $2$ discrete vacua and a combined continuum of vacua appear in this scenario. Once again, despite unclear physical implications, this case marks a clear transition between two distinct regimes.
 	\item When $\sigma>\frac{3}{2}$ the continuum of vacua  disappears. Nine extra vacua appear, including a vacuum point inside $S$ (point $13$). This increases the total number of vacua to $13$.
 \end{itemize}
   Notice that both the number of vacua and their position in the plane differ from those in the example presented in Section $2$. 
 
\begin{figure}[h!]
	\centering
	\includegraphics[height=4.5cm]{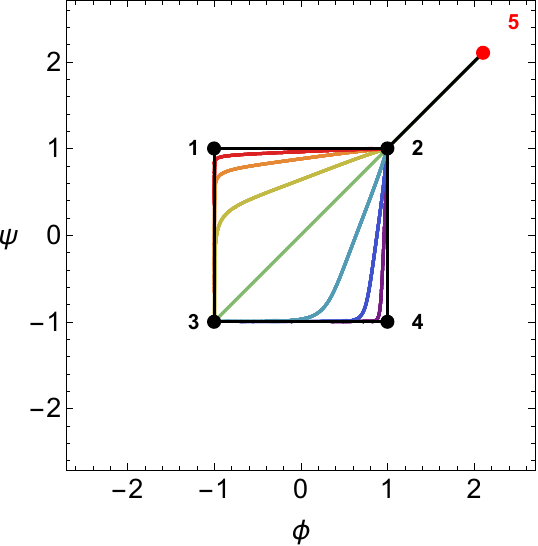}	\hspace{0.3cm}
	\includegraphics[height=4.5cm]{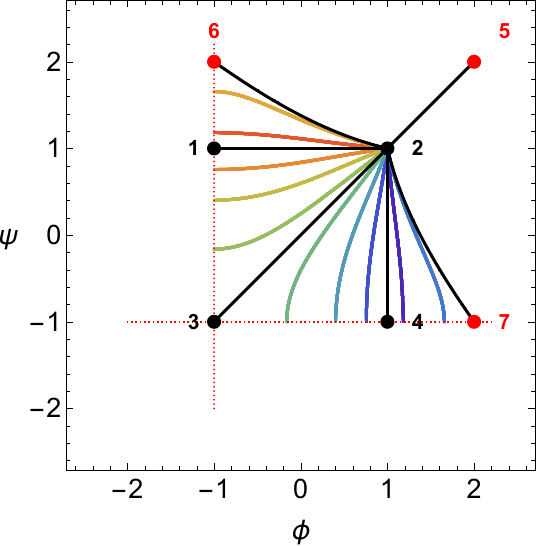} \hspace{0.3cm}
	\includegraphics[height=4.5cm]{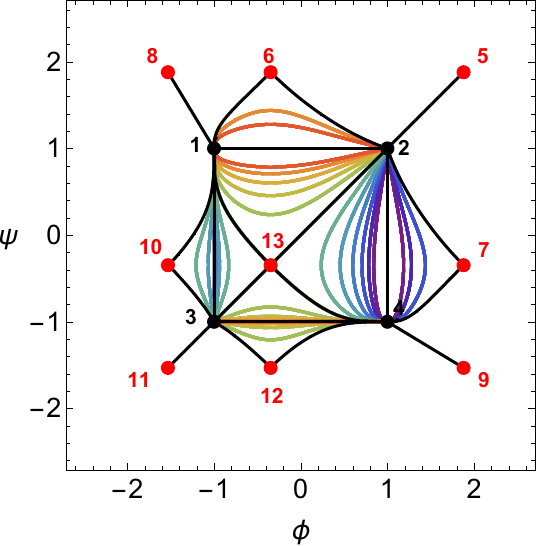}\\
	\includegraphics[height=4.5cm]{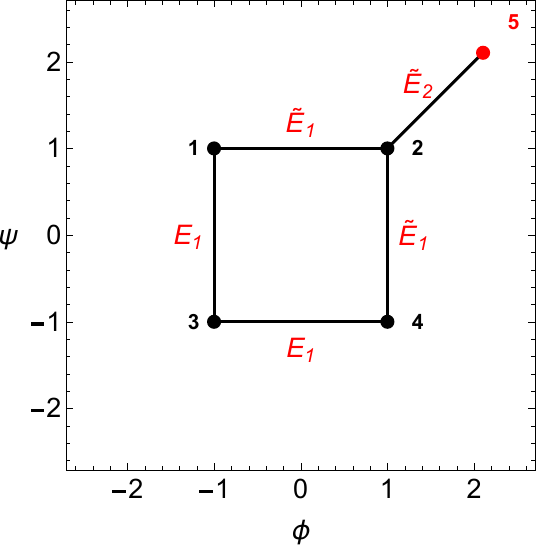}	\hspace{0.3cm}
	\includegraphics[height=4.5cm]{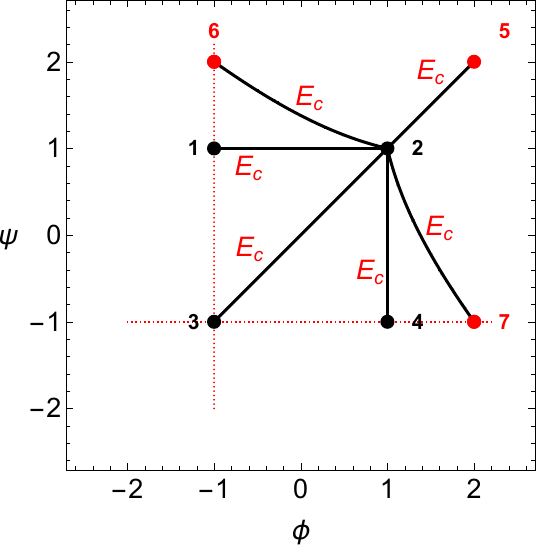} \hspace{0.3cm}
	\includegraphics[height=4.5cm]{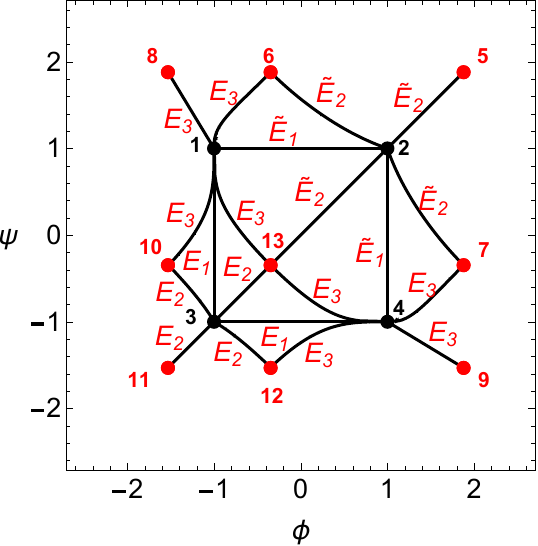}\\	
	\caption{\small Contourplot of the orbits of the emerging family of kinks (top panels) for different values of the constant $C$ for $\sigma=1$ (left), $\sigma=\frac{3}{2}$ (middle) and $\sigma=3$ (right) and energies \eqref{eq:EnergiesPhi4Double} of boundary and isolated kinks (bottom panels). The four inherited vacua are represented in black while the extra ones are depicted in red. Boundary and isolated kinks are represented in black.}
	\label{fig:CompositePhi4B}
\end{figure}

 This superpotential produces the following pair of Bogomol'nyi equations
$$\frac{d\phi}{dx}= \pm \left[1+\sigma\left(\psi-\frac{\psi^3}{3}\right)\right] ~ (\phi^2-1) \, , \qquad \frac{d\psi}{dx}= \pm \left[1+\sigma \left(\phi-\frac{\phi^3}{3}\right)\right] ~ (\psi^2-1) \, ,$$
which once more cannot be directly integrated. However, kinks corresponding to the trial orbits $\phi=\pm 1$ or $\psi=\pm 1$ can be identified. Indeed, these boundary kinks correspond to solutions of the $\phi^4-$model through the reparametrization \eqref{eq:ReparametrisationGeneral}
\begin{equation*}
	\frac{d\phi}{d\tilde{x}}= \pm ~ (\phi^2-1) \qquad \text{or} \qquad \frac{d\psi}{d\tilde{x}}= \pm ~ (\psi^2-1)\, . 
\end{equation*}
except for the case when $\sigma=\frac{3}{2}$, when Bogomol'nyi equations hold trivially when $\phi=-1$ or $\psi=-1$ respectively. Let us again denote as $\Phi_{i,j}$ the boundary or isolated kink that asymptotically connects the vacua $i$ and $j$. Now, from Bogomol'nyi equations the orbit flow equation \eqref{eq:OrbitFlowEqsMixing} can be derived and integrated, which reads
\begin{equation}\label{eq:ch6OrbitCompositeDoublephi4}
	\left|\frac{1-\phi}{1-\psi}\right|^{2\sigma+3} \left|\frac{1+\phi}{1+\psi}\right|^{2\sigma-3}=C \, e^{\sigma \left(\phi^2-\psi^2\right)}\, ,
\end{equation}
where $C>0$ is a constant that distinguishes between members of the arising family of kinks, see Figure \ref{fig:CompositePhi4B}. Let us once again denote as $\Sigma_{i,j}(C)$ the arising family of kinks that joins vacua $i$ and $j$.

It is easy to check that when $\sigma\neq \frac{3}{2}$ the critical points of the superpotential \eqref{eq:SuperpotentialDoublePhi4}, which are the arising vacua of the model, are non-degenerate and that all non-empty intersection of stable and unstable manifolds is transverse since no solution between saddle points can exist. This allows us to exclude the presence of families of kinks between specific points. In this context, a kink that asymptotically link critical points $p$ and $q$ can be found when the dimension of the intersection of the unstable $\mathcal{W}^{u}(p)$ and the stable $\mathcal{W}^{s}(q)$ manifolds
\begin{equation}\label{eq:DimensionPrerequisite}
	{\rm dim} \, \left(\mathcal{W}^{u}(p)\cap \mathcal{W}^{u}(q)\right)={\rm dim} \, \mathcal{W}^{u}(p) +{\rm dim} \, \mathcal{W}^{s}(q) -2\,,
\end{equation}
is at least one \cite{Smale1961,Bott1982}. One-parameter families of solutions, on the other hand, require that this dimension is at least two. A dimension check \eqref{eq:DimensionPrerequisite} reveals that isolated kinks between a saddle point and an inherited minima or maxima may be found. However, it also follows that one-parameter families of kinks must link inherited minima and maxima. This is reflected in the two different ``regimes'' that emerge depending on the strength of the coupling $\sigma$. In these regimes, as well as during the transition, the swapping symmetry $V(\phi,\psi;\sigma)=V(\psi,\phi;\sigma)$ of the potential \eqref{eq:PotentialExtensionPhi4} becomes apparent: 
\begin{itemize}
	\item When $0<\sigma<\frac{3}{2}$ four boundary kinks $\Phi_{1,2}$, $\Phi_{2,4}$, $\Phi_{3,4}$ and $\Phi_{1,3}$, which correspond to $\phi^4-$model solutions, emerge forming the edges of the square $S$. Moreover, the emergence of the extra vacuum point $5$ causes an extra isolated kink $\Phi_{1,5}$ to appear. Note that the behavior inside $S$ resembles that of the double $\phi^4-$model. However, instead of two families of kinks that asymptotically join diagonally aligned vacua, as occurs in the conventional double $\phi^4-$model, only one family of kinks $\Sigma_{2,3}(C)$ is obtained. Indeed, the analogue mechanical model is not separable for $\sigma\neq 0$. Moreover, their trajectories are asymmetric in $x$ because the form of the potential around vacua $(1,1)$ and $(-1,-1)$ is different, see Figure \ref{fig:CompositePhi4bVacuaPotential}. Notice also that limits of the constant $C\rightarrow\infty$ and $C\rightarrow 0$ correspond to the combination of boundary kinks $\Phi_{1,3}$ and $\Phi_{1,2}$ and of boundary kinks $\Phi_{3,4}$ and $\Phi_{2,4}$ respectively. 
	\item When $\sigma=\frac{3}{2}$ a continuum of vacua emerges, including points $1$, $3$, $4$, $6$ and $7$ in it. This scenario produces kinks joining different points of the continuum of vacua to point $2$ and an isolated kink $\Phi_{2,5}$ from point $2$ to the extra vacua $5$. Note that $\Phi_{2,6}$ and $\Phi_{2,7}$ correspond to the minimum value of $C$ for which finite energy solutions exist outside $S$. As the value of $C$ increases, solutions tend to $\Phi_{1,2}$ and $\Phi_{2,4}$, connecting to solutions inside $S$. Once more, the vanishing reparametrization factor corresponds to the impossibility of having solutions inside de continuum of vacua. In this model, the value $\sigma=\frac{3}{2}$ reveals a transition in behavior that results in the disappearance of the family of kinks $\Sigma_{2,3}(C)$ and the isolation of every kink. Since the continuum of vacua are degenerate critical points of the superpotential, the dimension prerequisite \eqref{eq:DimensionPrerequisite} cannot be directly applied.  
	
	\item Lastly, for $\sigma>\frac{3}{2}$ the lost boundary kinks $\Phi_{1,3}$ and $\Phi_{3,4}$ are once again recovered. Furthermore, the extra vacua produce four additional isolated kinks $\Phi_{1,8}$, $\Phi_{2,5}$, $\Phi_{3,11}$ and $\Phi_{4,9}$, eight boundary kinks $\Phi_{1,6}$, $\Phi_{1,10}$, $\Phi_{2,6}$, $\Phi_{2,7}$, $\Phi_{3,10}$, $\Phi_{3,12}$, $\Phi_{4,7}$ and $\Phi_{4,12}$, outside $S$, and four inside  $\Phi_{1,13}$, $\Phi_{2,13}$, $\Phi_{3,13}$ and $\Phi_{4,13}$. Note how the presence of the vacuum point $13$ changes the behavior inside the square $S$, where the four boundary kinks divide $S$ into four regions. In each of these four regions one family of kinks can be identified $\Sigma_{1,2}(C)$, $\Sigma_{1,3}(C)$, $\Sigma_{2,4}(C)$, $\Sigma_{3,4}(C)$, which also extend outside the square. It is easy to check that, after the transition at $\sigma=\frac{3}{2}$, all the inherited vacua become either maxima or minima of the superpotential, allowing more than one family of kinks to arise \eqref{eq:DimensionPrerequisite}. It is worth highlighting that the kink variety of this model for $\sigma>\frac{3}{2}$ resembles that of the model presented in Section $2$ when $\sigma>\frac{3}{2}$ once the picture is rotated $\frac{\pi}{4}$. 
\end{itemize}
  
Even though the kink variety is similar to that arising in the example in Section $2$, instead of horizontal reflections, symmetries are reflections respect to the diagonal of the square, which also affects the distribution of energies. Indeed, in this model the energies of boundary and isolated kinks, see Figure \ref{fig:CompositePhi4Bvac}, are given now by the following labels 
\begin{equation}\label{eq:EnergiesPhi4Double}
	E_1=\left|\frac{4}{3}-\frac{8 \sigma}{9}\right|\,, \, \widetilde{E}_1=\left|\frac{4}{3}+\frac{8 \sigma}{9}\right|\,, \, E_2=\frac{\left(2\sigma-3\right)^2}{9 \sigma}\,, \, \widetilde{E}_2=\frac{\left(2\sigma+3\right)^2}{9 \sigma}\,, E_3=\left|\frac{4 \sigma}{9}-\frac{1}{\sigma}\right|\,, \, E_c=\frac{8}{3}\,.
\end{equation}
Once more, since the superpotential value is the same at any extra vacua $W(v_{extra})=-\frac{1}{\sigma}$, all kinks joining an extra vacuum to a given inherited vacuum have the same energy. In order to discuss the energy of members of families of kinks, let us examine each scenario separately:
\begin{itemize}
	\item When $0<\sigma<\frac{3}{2}$ the energies of the boundary and isolated kinks are:
	\begin{equation}
		E[\Phi_{1,2}]=E[\Phi_{2,4}]=\widetilde{E}_1\,, \qquad E[\Phi_{1,3}]=E[\Phi_{3,4}]=E_1\,, \qquad E[\Phi_{2,5}]=\widetilde{E}_2\,.
	\end{equation}The energy of members of the family of kinks $\Sigma_{2,3}(C)$ in this case  is independent of the value of $\sigma$. Indeed, the swapping symmetry $W(\phi,\psi)=W(\psi,\phi)$ implies that \eqref{eq:ConditionSameEnergy} vanishes for vacua $2$ and $3$.  It follows that the energy of members of this family of kinks is that of the standard double $\phi^4-$model:
	\begin{equation}
		E\left[\Sigma_{2,3}\right]=E[\Phi_{1,2}]+E[\Phi_{1,3}]=E[\Phi_{2,4}]+E[\Phi_{3,4}]=E_c \,,
	\end{equation} 
	 where the energy sum rules have been included. Indeed, one can easily check the relation $E_1+\widetilde{E}_1=E_c$ $\forall \sigma\in[0,\frac{3}{2}]$. 
	\item When $\sigma=\frac{3}{2}$ all the emerging isolated kinks share the same energy $E[\Phi_{i,j}]=E_c$. Indeed, the value of the superpotential is the same at any vacua, except for the vacuum $2$.
	\item  When $\sigma>\frac{3}{2}$ the energy of boundary and isolated kinks is
	\begin{align*}
		&E[\Phi_{1,2}]=E[\Phi_{2,4}]=\widetilde{E}_1\,, \qquad E[\Phi_{1,3}]=E[\Phi_{3,4}]=E_1\,, \\
		 &E[\Phi_{2,5}]=E[\Phi_{2,6}]=E[\Phi_{2,7}]=E[\Phi_{2,13}]=\widetilde{E}_2 \,, \quad E[\Phi_{3,10}]=E[\Phi_{3,11}]=E[\Phi_{3,12}]=E[\Phi_{3,13}]=E_2\,,\\
		 &E[\Phi_{1,6}]=E[\Phi_{1,8}]=E[\Phi_{1,10}]=E[\Phi_{1,13}]=E[\Phi_{4,7}]=E[\Phi_{4,9}]=E[\Phi_{4,12}]=E[\Phi_{4,13}]=E_3\,.		 
	\end{align*}
	 On the other hand, the energy of members of each family is given by 
	\begin{align*}
		&E\left[\Sigma_{1,2}\right]=E[\Phi_{1,2}]=E[\Phi_{1,6}]+E[\Phi_{2,6}]=E[\Phi_{1,13}]+E[\Phi_{2,13}]=\widetilde{E}_1\,,\\
		&E\left[\Sigma_{1,3}\right]=E[\Phi_{1,3}]=E[\Phi_{1,10}]+E[\Phi_{3,10}]=E[\Phi_{1,13}]+E[\Phi_{3,13}]=E_1\,,\\
		&E\left[\Sigma_{2,4}\right]=E[\Phi_{2,4}]=E[\Phi_{2,7}]+E[\Phi_{4,7}]=E[\Phi_{2,13}]+E[\Phi_{4,13}]=\widetilde{E}_1\,,\\
		&E\left[\Sigma_{3,4}\right]=E[\Phi_{3,4}]=E[\Phi_{3,12}]+E[\Phi_{4,12}]=E[\Phi_{3,13}]+E[\Phi_{4,13}]=E_1\,,
	\end{align*}
	where once more the different energy sum rules have been included. It is worth highlighting that this is consistent with the relation between different models $(\phi,\psi,-\sigma)\rightarrow(-\phi,-\psi,\sigma)$, which transforms $E_1\leftrightarrow \widetilde{E}_1$, $E_2\leftrightarrow \widetilde{E}_2$, $E_3\leftrightarrow E_3$. 	 
\end{itemize}

Note that this model extends the well-known double $\phi^4-$model, which is recovered when $\sigma=0$. However, as a result of the introduction of this higher order coupling terms, richer kink varieties emerge, especially when the critical value $\sigma=\frac{3}{2}$ is surpassed. In fact, the behavior of this model when $\sigma>\frac{3}{2}$ is similar to that of that of the previous section, even though the behavior for low $\sigma$ differs significantly.

\subsection{Composite sine-Gordon model with three fields}

In previous sections extensions of the $\phi^4-$model are constructed. In this section, the goal is to generalize the sine-Gordon model with three fields. This will allow us to identify more interesting kink varieties and add internal structure. Since the sine-Gordon model is integrable, the high-order terms introduced when $\sigma\neq 0$ not only will couple fields, but also shall break the integrability of the model. In order to ensure that potentials have discrete vacua, the superpotentials $W_A(\phi)$ and $W_B(\psi)$ are assumed to be non-negative functions whose sets of zeros are discrete. In particular, in this construction of an extended sine-Gordon model with three fields with analytical solutions, the two non-negative superpotentials that correspond to sine-Gordon models shall be chosen as:
\begin{equation*}
	W_A(\phi_1,\phi_2)=2-\sin{\phi_1}-\sin{\phi_2} \, , \qquad W_B(\psi)=1-\sin{\psi} \, .
\end{equation*}
  This allows the present procedure with $\alpha=\beta=1$ to engender a model in $\mathbb{R}^3$ with a positive semidefinite superpotential
 \begin{equation*}
 	W(\phi_1,\phi_2,\psi;\sigma)=W_A(\phi_1,\phi_2)+W_B(\psi) + \sigma \, W_A(\phi_1,\phi_2) \, W_B(\psi)\, ,
 \end{equation*}
 assuming that the parameter is also defined as non-negative $\sigma\geq 0$. This implies that the family of potential functions \eqref{eq:ch6potentialComposite} constructed from this superpotential 
 \begin{align}
 	V(\phi_1,\phi_2,\psi;\sigma)=&\frac{1}{2} \left[1+ \sigma \, \left(1-\sin{\psi}\right)\right]^2\left[\cos^2\phi_1+\cos^2\phi_2\right]\label{eq:PotentialSGC}\\
 	&+ \frac{1}{2} \left[1+ \sigma \, \left(2-\sin{\phi_1}-\sin{\phi_2}\right)\right]^2 \cos^2\psi \, ,\nonumber
 \end{align}
   presents as vacua only the Cartesian product of the vacua of the original models, regardless of the value of the parameter $\sigma$. In particular, the inherited vacua from the original sine-Gordon models are located at points
 \begin{equation*}
 	\mathcal{M}=\left\lbrace v_{n_1,n_2,n_3}=\left(\frac{\pi}{2}+n_1 \pi, \frac{\pi}{2}+n_2 \pi, \frac{\pi}{2}+n_3 \pi\right) \right\rbrace\, ,
 \end{equation*}
 for integers $n_1,n_2$ and $n_3$. In other words, the target space $\mathbb{R}^3$ is divided into cubes whose vertices correspond to the vacua $v_{n_1,n_2,n_3}$. Due to the periodicity of the potential \eqref{eq:PotentialSGC}, kinks are replicated across the lattice. Therefore, it suffices to focus on the cube defined by $\phi_1,\phi_2,\psi\in(-\frac{\pi}{2},\frac{\pi}{2})$, with vacua numbered from $1$ to $8$, see Figure \ref{fig:cubocarasSG}. On the other hand, the coupling between fields in the potential gives rise to the following Bogomol'nyi equations
  \begin{align}
 	\frac{d \phi_1}{dx}&=\mp \left[1+ \sigma \, \left(1-\cos\psi\right)\right] \cos\phi_1 \, , \label{eq:SineGordonBPS1}\\
 	\frac{d \phi_2}{dx}&=\mp \left[1+ \sigma \, \left(1-\cos\psi\right)\right] \cos\phi_2 \, ,\label{eq:SineGordonBPS2}\\
 	\frac{d \psi}{dx}&=\mp \left[1+ \sigma \, \left(2-\cos\phi_1-\cos\phi_2\right)\right] \cos\psi \, . \label{eq:SineGordonBPS3}	 
 \end{align}
 Similarly to the preceding examples, apart from the full-dimensional flow solutions, special solutions of these equations can be obtained by prescribing trial orbits and integrating the associated orbit equations. Let us distinguish between three types of solutions:
 \begin{itemize}
 	\item {\bf Boundary $\Phi_{i,j}$ kinks on the edges of the cube}: When two fields are fixed at values $\frac{\pi}{2}$ or $-\frac{\pi}{2}$, boundary kinks emerge as edges of the cube. Indeed, Bogomoln'yi equations, \eqref{eq:SineGordonBPS1}, \eqref{eq:SineGordonBPS2} and \eqref{eq:SineGordonBPS3}, reduce for the unfixed field to kinks of the well-known sine-Gordon model, up to a reparametrization in the spatial line. These are depicted in dashed blue lines in Figure \ref{fig:cubocarasSG}. Similarly to previous sections, the boundary kinks asymptotically joining the adjacent vacua $i$ and $j$, will be denoted as $\Phi_{i,j}$. 
 	
 	\item {\bf Families of kinks $\Psi_{k,l}(\gamma)$ on the cube faces}: If only one of the coordinates is fixed $\phi_j=\pm\frac{\pi}{2}$, the integration of the orbit equations leads to families of kinks on the faces of the cube, see Figure \ref{fig:cubocarasSG}. The orbit equations can be integrated:
 	\begin{itemize}
 		\item When $\phi_1=\pm \frac{\pi}{2}$ or $\phi_2=\pm \frac{\pi}{2}$ a family of orbit equations of label $\gamma_i\in\mathbb{R}$ is obtained: 
 		\begin{equation}
 			(1+2\sigma) \arctan\left(\sin \phi_i\right)-(1+\sigma) \arctan\left(\sin \psi\right)=\sigma \, (\phi_i-\psi)+ \gamma_i\,, \qquad i=1,2\,.
 		\end{equation}
 		\item When $\psi=\pm \frac{\pi}{2}$ one obtains a solution that can be written in terms of the Gudermannian function, denoted as ${\rm Gd}$: 
 		\begin{equation}\label{eq:GudermannianOrbit}
 			\phi_1={\rm Gd} \, \left[C+ {\rm Gd}^{-1} \, \left[\phi_2\right]\right]\, ,
 		\end{equation}
 		and where and $C\in\mathbb{R}$ is an integration constant that labels each member of the family of kinks. For the sake of simplicity this equation has only been solved for the cube defined by $\phi_1,\phi_2,\psi\in(-\frac{\pi}{2},\frac{\pi}{2})$, even though this is straightforwardly generalizable to other cubes. Let us denote as $\Psi_{k,l}$ the families of kinks on the faces of the cubes whose boundaries are the $\Phi_{i,j}$ kinks.
 	\end{itemize}
 	
  	\begin{figure}[h!]
 	\centering
 	\includegraphics[height=3cm]{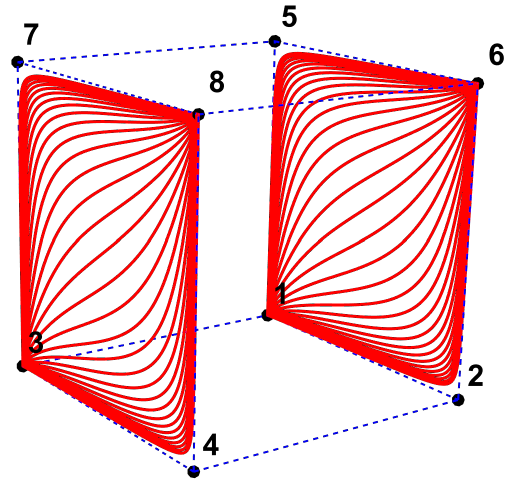}	\hspace{1cm}
 	\includegraphics[height=3cm]{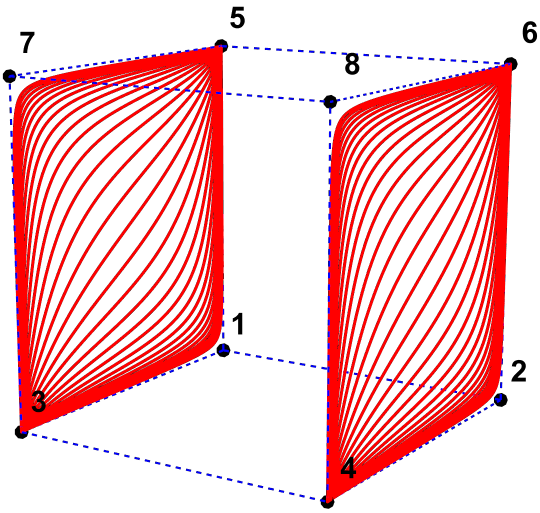}\hspace{1cm}
 	\includegraphics[height=3cm]{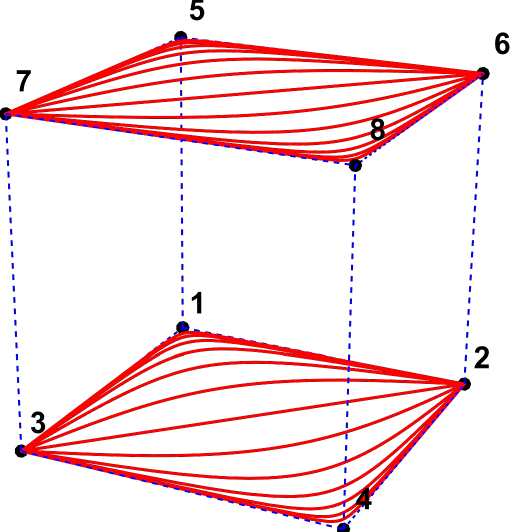}\\
 	\vspace{0.3cm}\hspace{-0.2cm}
 	\includegraphics[height=3cm]{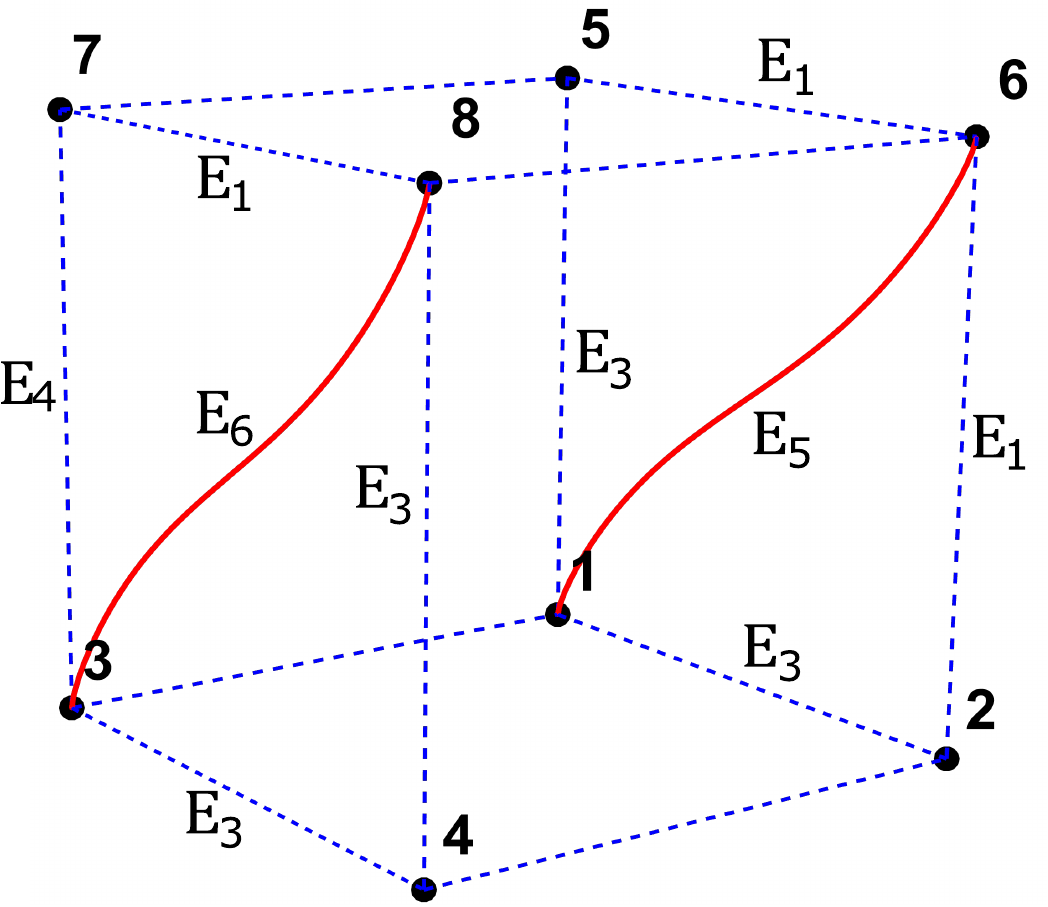}	\hspace{1cm}
 	\includegraphics[height=3cm]{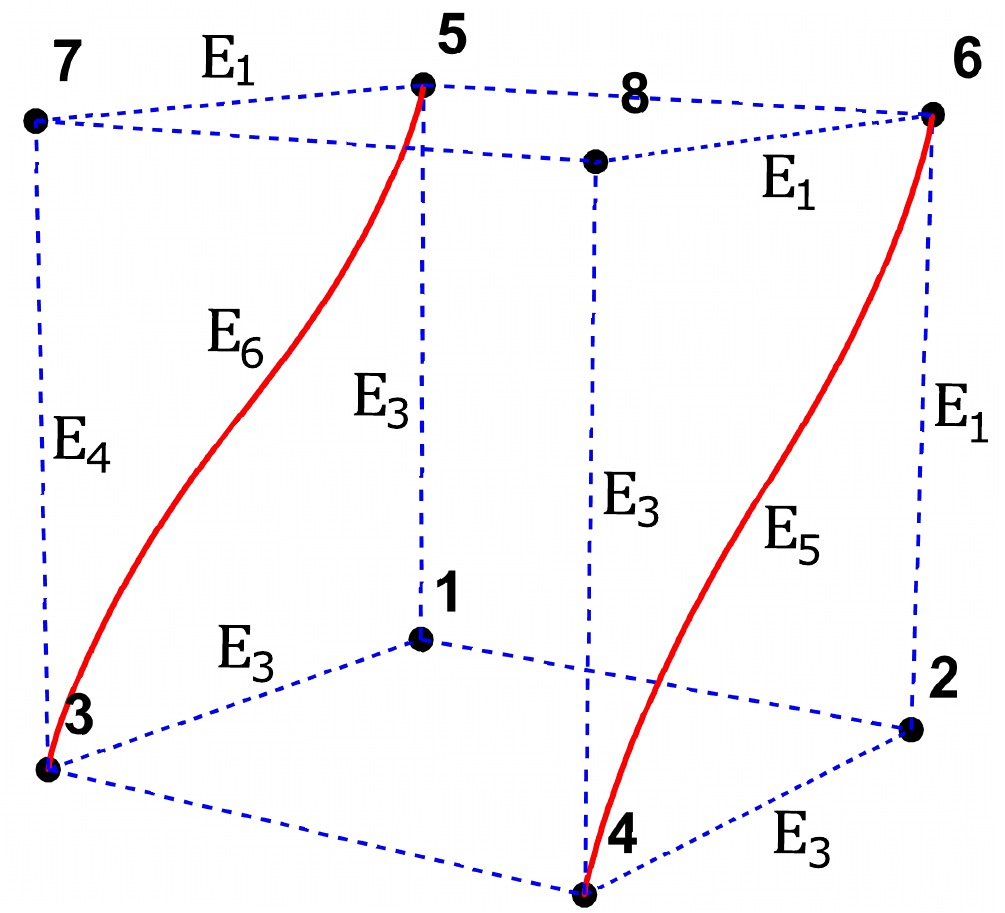}\hspace{1cm}
 	\includegraphics[height=3cm]{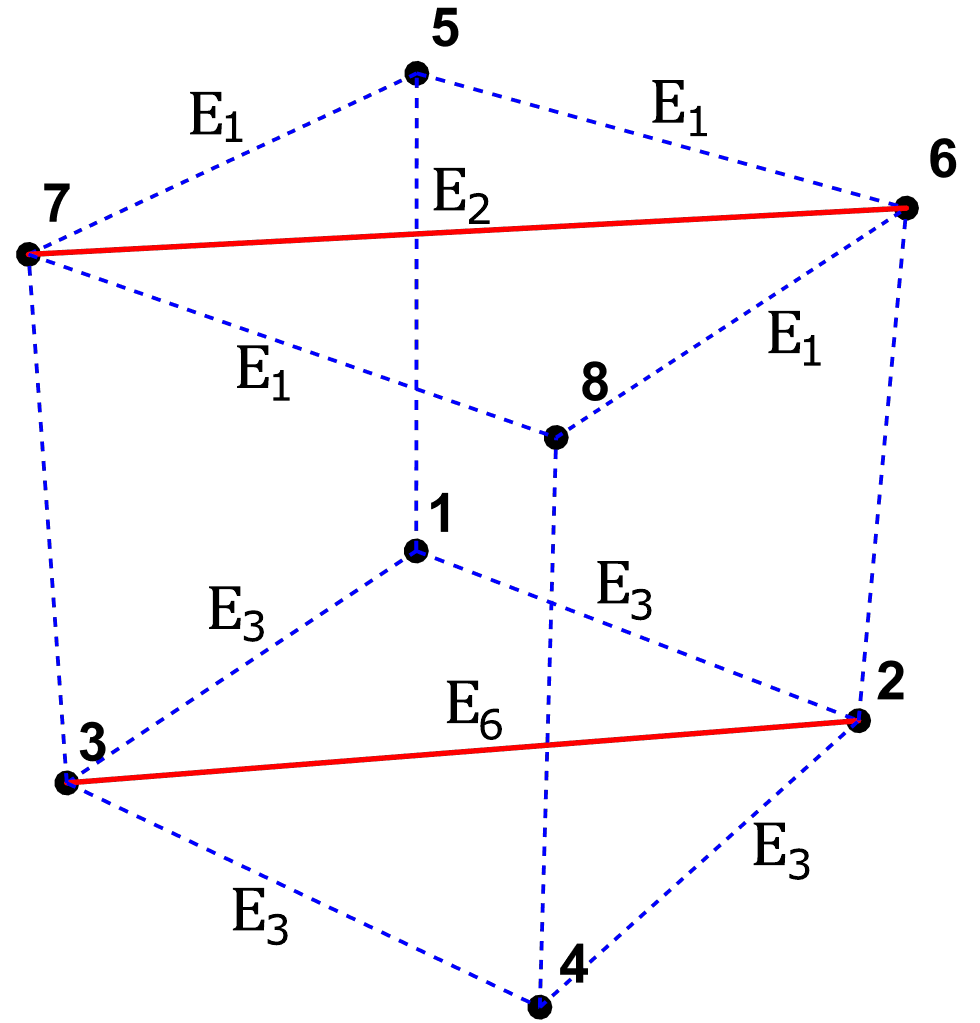} 			
 	\caption{\small Members of the families of kinks $\Psi_{1,6}$ and $\Psi_{3,8}$ (left panel), $\Psi_{3,5}$ and $\Psi_{4,6}$ (middle panel) and $\Psi_{2,3}$ and $\Psi_{6,7}$ (right panel) are shown for $\sigma=0.5$ and various values of the corresponding integration constants $C$, $\gamma_1$ or $\gamma_2$ (top panels). The bottom panels display the corresponding energies. $\Phi_{i,j}$ kinks have been depicted in blue dashed lines forming the edges of the cube.}
 	\label{fig:cubocarasSG}
 \end{figure}

 \item {\bf Family of kinks $\Sigma_{3,6}(\gamma,C)$ in the interior of the cube}:  Consider the orbits of the general solution with no constant fields 
  \begin{align}
 	\frac{d\phi_1}{d\phi_2}=& \frac{\cos\phi_1}{\cos\phi_2} \, ,\label{eq:Orbits3DA}\\
 	\frac{d\phi_2}{d\psi}=&\frac{1+ \sigma\left(1-\cos\psi\right)}{1+ \sigma\left(2-\cos\phi_1-\cos\phi_2\right)}\frac{\cos\phi_2}{\cos\psi} \, .\label{eq:Orbits3DB}
 \end{align}
 Solutions of equation \eqref{eq:Orbits3DA} provides the first orbit equation, given once more by \eqref{eq:GudermannianOrbit} with integration constant $C$. Moreover, this orbit equation can be substituted in \eqref{eq:Orbits3DB}, leading to the second orbit equation
	\begin{equation}
		(1+2\sigma) \arctan\left(\sin \phi_i\right)-(1+\sigma) \arctan\left(\sin \psi\right)+F(\phi_2,\psi;C)=\sigma \, (\phi_i-\psi)+ \gamma\,,
	\end{equation}
 	where $\gamma\in(0,\infty)$ is an integration constant and $F(\phi_2,\psi;C)$ is a function defined as follows
 		\begin{equation}
 		F(\phi_2,\psi;C)=-2\sigma \arctan\left(\sinh C + \cosh C \tan\frac{\phi_2}{2}\right)\,.			
 	\end{equation}
 	As the parameter $\sigma$ increases in value, trajectories are affected more strongly by the coupling terms, see Figure \ref{fig:cubofamiliasSG}. All the depicted solutions share the same value of $C$ forming a surface inside the cube. All possible values of this constant $C$ will produce surfaces that densely fill the interior of the cube. Let us denote this family of kinks between vacua $3$ and $6$ as $\Sigma_{3,6}(\gamma,C)$. 
  	 	\begin{figure}[h!]
 	\centering
 	\includegraphics[height=4cm]{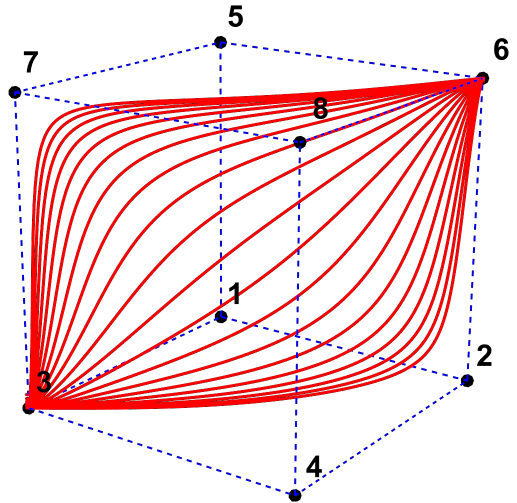}	\hspace{0.5cm}
 	\includegraphics[height=4cm]{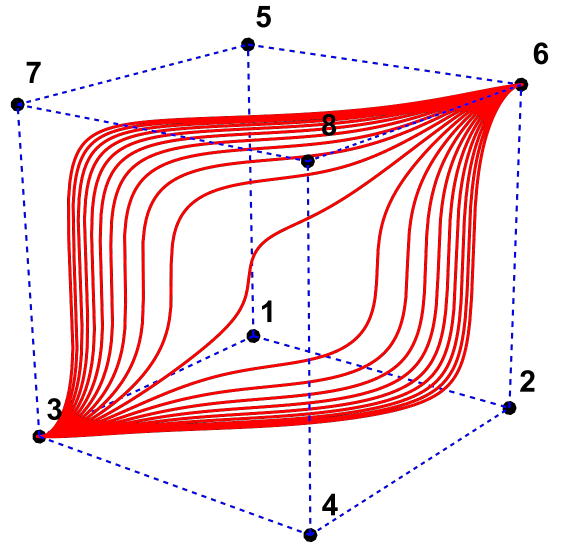}			
 	\caption{\small Different members of the families of kinks $\Sigma_{3,6}$ are shown for $C=0.5$ and various values of $\gamma$ when $\sigma=\frac{1}{2}$ (left) and $\sigma=100$ (right). As $\sigma$ increases, the kink orbits curve more sharply around the center of the cube. $\Phi_{i,j}$ kinks have been depicted in blue dashed lines forming the edges of the cube.}
 	\label{fig:cubofamiliasSG}
 \end{figure}
\end{itemize}
 \noindent Once more, the kink energies can be computed explicitly. For the $\Phi_{i,j}$ kinks, three different energy levels, $E_1$, $E_3$ and $E_4$, appear
 \begin{align*}
	&E[\Phi_{1,2}]=E[\Phi_{1,3}]=E[\Phi_{1,5}]=E[\Phi_{2,4}]=E[\Phi_{3,4}]=E[\Phi_{4,8}]=E_3\,, \qquad E[\Phi_{3,7}]=E_4\,,\\
	&E[\Phi_{2,6}]=E[\Phi_{5,6}]=E[\Phi_{5,7}]=E[\Phi_{6,8}]=E[\Phi_{7,8}]=E_1\,,		 
\end{align*}
 while other three different values of energy, $E_2$, $E_5$ and $E_6$, appear for $\Psi_{k,l}$ kinks
  	\begin{align*}
 	&E\left[\Psi_{1,6}\right]=E[\Phi_{1,2}]+E[\Phi_{2,6}]=E[\Phi_{1,5}]+E[\Phi_{5,6}]=E_5\,,\\
 	&E\left[\Psi_{2,3}\right]=E[\Phi_{1,2}]+E[\Phi_{1,3}]=E[\Phi_{2,4}]+E[\Phi_{3,4}]=E_6\,,\\
 	&E\left[\Psi_{3,5}\right]=E[\Phi_{1,3}]+E[\Phi_{1,5}]=E[\Phi_{3,7}]+E[\Phi_{5,7}]=E_6\,,\\
 	&E\left[\Psi_{3,8}\right]=E[\Phi_{3,4}]+E[\Phi_{4,8}]=E[\Phi_{3,7}]+E[\Phi_{7,8}]=E_6\,,\\
 	&E\left[\Psi_{4,6}\right]=E[\Phi_{2,4}]+E[\Phi_{2,4}]=E[\Phi_{4,8}]+E[\Phi_{6,8}]=E_5\,,\\
 	&E\left[\Psi_{6,7}\right]=E[\Phi_{5,6}]+E[\Phi_{5,7}]=E[\Phi_{6,8}]+E[\Phi_{7,8}]=E_2\,,
 \end{align*}
 where each label corresponds to
 \begin{align*}
 	E_1&=2\,,\quad E_2=4\,, \quad E_3=2+4\sigma\,,\quad E_4=2+8\sigma\,,\quad E_5=4+4\sigma\,,\quad E_6=4+8\sigma\,.
 \end{align*}
  see Figure \ref{fig:cubocarasSG}. Once again, the energy relations incorporate the energy sum rules, which reflect the limits of families of kinks. For each face of the cube, taking the appropriate limits of the constants $C$ and $\gamma$ causes the corresponding family of $\Psi_{k,l}$ kinks to approach two specific combinations of $\Phi_{i,j}$ kinks. Lastly, the energy of members of the family of kinks $\Sigma_{3,6}(C,\gamma)$, which fill the interior of the cube, is $E_{in}=6+8 \sigma$. Notice that this is the energy of members of the family of kinks of the three-field sine-Gordon model, with an additional contribution $8 \sigma$ whose size is modulated by the strength of the coupling. Finally, energy sum rules relating $\Sigma_{3,6}(C,\gamma)$ and $\Psi_{k,l}$ and $\Phi_{i,j}$ kinks can be identified
  \begin{equation}
  	E\left[\Sigma_{3,6}\right]=E[\Psi_{2,3}]+E[\Phi_{2,6}]=E[\Psi_{3,5}]+E[\Phi_{5,6}]=E[\Psi_{3,8}]+E[\Phi_{6,8}]=E_{in}\,,
  \end{equation}
  where all the previously described energy sum rules also apply.
  
  Therefore, this model represents a generalization of the sine-Gordon model with three fields, with a significantly richer kink variety modulated by the coupling parameter $\sigma$.

\section{Conclusions}

In this work a very particular field theory with a polynomial potential with high-order terms is studied. This has allowed us to identify the properties that allows its analytical solutions to be found, which leads to a formalism to construct new field theories from others while maintaining certain analytical information. One-parameter families of superpotentials on the new target space, the Cartesian product of the original target spaces, are defined so that the dynamics of the originally isolated field theories are entangled in a larger coupled field theory. This allows the model to inherit vacua and kinks from the original models, but also the appearance of extra ones. 

For these composite models, the Hessian of the new superpotential at inherited vacua maintains the Hessian's eigenvalues of the original models up to some constants. These constants, however, may change the stability of these solutions between maxima, minima, saddle and degenerate points. When extra vacua appear, these always become saddle points of the superpotential, where the value of the superpotential is the same fixed value, with only two non-vanishing eigenvalues. Hyperplanes, defined by the components of the inherited vacua, split the target space into subspaces where at most one extra vacuum can be found. In the boundary of these regions between these hyperplanes, which solutions are not allowed to cross, boundary kinks are found. Inside these regions, isolated kinks and families of kinks are expected to arise. When the critical points of the superpotential are non-degenerate and the stable and unstable manifolds intersections are transverse, the existence of certain families of kinks can be excluded. Indeed, in the second example in $\mathbb{R}^2$, where these conditions hold for $\sigma\neq \frac{3}{2}$, families of kinks can only arise between inherited minima and maxima.

    This mechanism of construction of composite models, firstly introduced as an example where two $\phi^4-$models are combined, is also employed in two other examples with two and three fields, where the analytical derivation of the kink orbits is guaranteed. The first composite model, presented as a model to study in Section $2$, can be understood for low values of the parameter $|\sigma|$, as the extension in the plane of a $\phi^4-$model with one field. However, after a critical value of the coupling parameter, the dynamics is dominated by the coupling between fields, which now exhibit a different behavior. The second example is a generalization of the $\phi^4-$model in the plane, recovering this model when $\sigma=0$. In these two scenarios, this mechanism allows us to introduce an extra number of vacua in the theory, which in general depends on the parameter of the coupling. It is also worth highlighting the fact that while these two models exhibit different behaviors for weak couplings, these resemble in the regime when the coupling dominates, even though the arising symmetries are not the same.   
Both models exhibit a ``transition'' for a certain value of the coupling parameter in which continua of vacua appear and boundary kinks disappear. Some of the vacua become degenerate at this transition. However, in the second model members of the family of kinks become ``isolated''. It is also worth highlighting that the relation between potentials under the transformation $\sigma \rightarrow -\sigma$ has allows us to reduce the study to non-negative coupling constants without loss of generality. Lastly, a generalization of the sine-Gordon field theory with three fields is constructed. This generalization allows us to introduce coupling terms in the polynomial potential and still being able to identify analytical solutions. This results in the appearance of an extra internal structure, once more modulated by the coupling constant $\sigma$. Furthermore, this extension is constructed so that no continuous vacua appear in the new field theory, inheriting only the Cartesian product of the original vacua in the respective sine-Gordon models. That is, it preserves the vacuum manifold of the standard non-coupled sine-Gordon model with three fields. 

On the other hand, the energies of all these kinks depend in general on the coupling parameter. For instance, in the case of the composite sine-Gordon model with three fields, the energy of all members of all families of kinks increases as the coupling parameter increases. BPS kinks corresponding to the new superpotential are compelled to connect at least one inherited vacuum. This also implies that the information about the energy of BPS kinks is contained in the parameters of the model $\alpha,\beta,\sigma$ as well as in the value of the superpotential at the inherited vacua. Nevertheless, superpotentials of the original models can be chosen such that the energy of the emerging family of kinks becomes independent of the particular model $\sigma$, as occurs in the example shown in Section $3$. Energy sum rules have been also found to depend on the coupling parameter, arising different rules for different regimes in both extension of the $\phi^4-$model.

It is worth highlighting that generalizing the double $\phi^4-$model and the sine-Gordon model with three fields while being able to identify analytically solutions could have significant applications in physics. These standard models are widely used to describe various phenomena, and their generalized versions could further extend their predictive capabilities. Lastly, this procedure can be generalized to the context of Sigma models, which is a line of research that shall be explored in the future.

\section{Acknowledgments}

This research was funded by the Spanish Ministerio de Ciencia e Innovación (MCIN) with funding from the European Union NextGenerationEU
(PRTRC17.I1) and the Consejería de Educación, Junta de Castilla y León, through QCAYLE project, as well as grant PID2023-148409NB-I00 MTM funded by MCIN/AEI/10.13039/501100011033.

\end{document}